\documentclass{jfm}
\usepackage[OT1]{fontenc} 
\usepackage{graphicx}
\usepackage{float}
\usepackage{epstopdf, epsfig}
\usepackage{subfigure}
\usepackage{color}
\usepackage{psfrag}
\usepackage{natbib}
\usepackage{multirow}
\usepackage{psfrag}
\usepackage{xfrac}
\usepackage{sidecap}
\usepackage{booktabs}

\usepackage{pstricks}
\usepackage{tikz}
\usepackage{amsmath}

\usepackage{calc}

\usepackage{caption}
\usepackage{rotating}
\usepackage{multirow}
\usepackage{tabularx}
\usepackage{upgreek}
\usepackage[normalem]{ulem}
\usepackage{siunitx}

\usepackage{bm}
\usepackage{mathrsfs}
\usepackage{bigdelim}
\usepackage{pifont}
\usepackage{txfonts} 

\definecolor{green_m}{rgb}{0,0.5,0}
\definecolor{magenta_m}{rgb}{1,0,1}
\definecolor{magenta1}{rgb}{.75,0,0.75}
\definecolor{GroupRe}{rgb}{0.76,0.05,0.53}
\definecolor{Groupks}{rgb}{0.23,0.56,1}

\DeclareRobustCommand\TriPinkOne{\includegraphics[trim = 1mm 1mm 0mm 0mm]{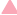}}
\DeclareRobustCommand\TriPinkTwo{\includegraphics[trim = 1mm 1mm 0mm 0mm]{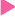}}
\DeclareRobustCommand\TriPinkFour{\includegraphics[trim = 1mm 1mm 0mm 0mm]{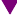}}

\DeclareRobustCommand\TriThree{\includegraphics[trim = 5mm 1mm 4mm 2mm]{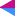}}

\DeclareRobustCommand\CircBlueOne{\includegraphics[trim = 1mm 1mm 0mm 0mm]{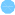}}
\DeclareRobustCommand\DmdBlueThree{\includegraphics[trim = 1mm 1mm 0mm 0mm]{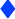}}

\begin{document}

\author{Mogeng Li\aff{1}, Charitha M. de Silva\aff{1, 2}, Daniel Chung\aff{1}, Dale I. Pullin\aff{3}, Ivan Marusic\aff{1} and Nicholas Hutchins\aff{1}}

\title{\color{black} Experimental study of a turbulent boundary layer with a rough-to-smooth change in surface conditions at high Reynolds numbers}

\shorttitle{Experimental study of a TBL with an R$\rightarrow$S change at high $Re$}
\shortauthor{M. Li and others}

\affiliation{\aff{1}Department of Mechanical Engineering, University of Melbourne, Victoria 3010, Australia
\aff{2}School of Mechanical and Manufacturing Engineering, University of New South Wales, NSW 2052, Australia
\aff{3}Graduate Aerospace Laboratories, California Institute of Technology, CA 91125, USA}

\date{Received: date / Accepted: date}

\maketitle

\begin{abstract}
This study presents an experimental dataset documenting the evolution of a turbulent boundary layer downstream of a rough-to-smooth surface transition. To investigate the effect of upstream flow conditions, two groups of experiments are conducted. For the \emph{Group-Re} cases, a nominally constant viscous-scaled equivalent sand grain roughness $k_{s0}^+\approx160$ is maintained  on the rough surface, while the friction Reynolds number $Re_{\tau 0}$ ranges from 7100 to 21000. For the \emph{Group-ks} cases, $Re_{\tau 0}\approx14000$ is maintained while $k_{s0}^+$ ranges from 111 to 228. The wall-shear stress on the downstream smooth surface is measured directly using oil-film interferometry to redress previously reported uncertainties in the skin-friction coefficient recovery trends. In the early development following the roughness transition, the flow in the internal layer is not in equilibrium with the wall-shear stress. This conflicts with the common practise of modelling the mean velocity profile as two log laws below and above the internal layer height, as first proposed by Elliott (\textit{Trans. Am. Geophys. Union}, vol. 39, 1958, pp 1048--1054). As a solution to this, the current data are used to model the recovering mean velocity semi-empirically by blending the corresponding rough-wall and smooth-wall profiles. The over-energised large-scale motions leave a strong footprint in the near-wall region of the energy spectrum, the frequency and magnitude of which exhibit dependence on $Re_{\tau 0}$ and $k_{s0}^+$ respectively. The energy distribution in near-wall small scales is mostly unaffected by the presence of the outer flow with rough-wall characteristics, which can be used as a surrogate measure to extract the local friction velocity.

\end{abstract}


\section{Introduction} \label{sec:intro}
\begin{figure}
    \centering
    \includegraphics[width=5in, clip]{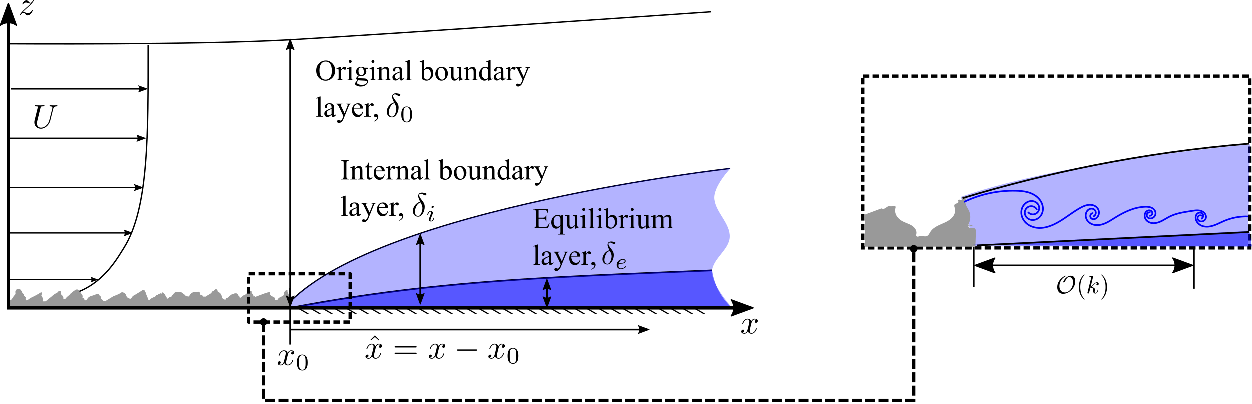}
    \caption{Schematic of a turbulent boundary layer over a rough-to-smooth change in surface condition. The roughness transition occurs at $x_0$, and $\hat{x} = x-x_0$ denotes the fetch downstream of the transition. The inset is a zoom-in view of the boxed region close to the rough-to-smooth change, where a trail of vortices shed from the roughness elements.}
    \label{fig:IBL_sketch}
\end{figure}

Surface roughness with heterogeneity is present in wall-bounded turbulent flows in a variety of conditions. Examples include the patchiness of biofouling on the hull of a ship or the changes in the surface roughness conditions that occur at the interface between forest and grasslands. Understanding the flow response to such a change in the surface condition is beneficial to various practical applications such as predicting the drag penalty introduced by non-uniform fouling or improving the weather forecast in regions with a change of terrain. Although roughness heterogeneity can occur in a number of configurations, here we consider a simple scenario, namely a sudden rough-to-smooth surface transition occurring in the streamwise direction, as depicted in figure \ref{fig:IBL_sketch}. Upstream of the transition, a turbulent boundary layer develops on a rough wall with equivalent sand grain roughness height $k_s$. $x$ is the streamwise direction, $x_0$ is the streamwise location of the surface transition and $\hat{x} \equiv x - x_0$ is the distance downstream of the transition. At $x=x_0$, the surface switches to a smooth wall, while the boundary layer continues to evolve and gradually adjusts to the new surface. The effect of the new surface condition is firstly felt in the near-wall region of the boundary layer and then gradually propagates to the interior of the flow \citep{Garratt1990}. The layer that separates the modified near-wall region from the unaffected oncoming flow further away from the wall is generally referred to as the internal boundary layer (IBL) with a thickness denoted by $\delta_i$. The complete adjustment of flow statistics to the downstream wall condition (referred to here as equilibrium) also first emerges in the region immediately adjacent to the smooth wall. This region is referred to as the equilibrium layer (EL) \citep{Garratt1990, savelyev2005internal} with thickness $\delta_e$. In most cases, a large portion of the flow within the IBL has still not fully adapted to the local wall condition \citep{antonia1972response, rouhi2018, ismail2018simulations, MogengJFM2019}, and a general consensus is that $\delta_e\approx0.05\delta_i$, where $\delta_i$ is defined based on the shear stress profile adjustment downstream of a rough-to-smooth change \citep[see][]{Rao1974,Shir1972}. The term `transition layer' is sometimes used for the region between the EL and IBL \citep{savelyev2005internal}. As shown in the inset of figure 1, immediately downstream of the rough-to-smooth change, a region exists where we would expect a recirculation region to form, as a result of the step in the surface elevation between the rough and smooth walls which produces vortex shedding from the roughness crests. Similar to backward facing steps \citep{kostas2002particle, barri2010dns, wu2013turbulent, itt2018}, this region (which we refer to as the `roughness trailing wake') persists over a streamwise fetch that scales on the roughness height, before the intensity attenuates and falls below the local turbulence intensity of the existing turbulent boundary layer.

The study on flows over a streamwise roughness transition started in the meteorology community driven by the need to understand the effect of a change in the terrain on the microclimate \citep{Elliott1958, Bradley1968, Shir1972, Rao1974}. These studies focused on deep surface layers with the upstream and downstream surfaces assumed to be fully rough. Such flows are dependent on a single parameter, the magnitude of roughness change (usually denoted by $M$ and defined as a function of the ratio between the upstream and downstream roughness lengths). Further studies on turbulent boundary layers with a finite thickness \citep[mainly experimental, see][for example]{antonia1972response, Hanson2016, MogengJFM2019} and channel flows \citep[mainly numerical, see][for example]{bou2004large, Saito2014, ismail2018simulations, rouhi2018} have appeared in recent years. In either case, a new length scale of the outer flow (the boundary layer thickness or the channel half height) becomes more relevant, and the flow response is expected to be dependent on both the magnitude of roughness change and a new non-dimensional parameter involving the outer length scale. This will be especially meaningful further away from the roughness transition where the IBL is comparable with the local boundary layer thickness.

Downstream of the rough-to-smooth transition, the skin-friction coefficient undershoots the corresponding smooth-wall value before gradually recovering in the far-field \citep{antonia1972response, Hanson2016, ismail2018simulations, rouhi2018, MogengJFM2019}. The deviation of the mean flow within the IBL from a self-similar canonical smooth-wall boundary layer leads to an underestimation of the skin-friction coefficient in both `indirect' measurements \citep{loureiro2010distribution, MogengJFM2019}, such as the Clauser-chart method or Preston tube \citep{Patel1965}, and the predictive model of \citet{Elliott1958}, where an equilibrium log-law profile is assumed within the IBL. There have been various attempts to model the adjusting flow within the IBL, either by using a local shear stress that varies in the wall-normal direction in a mixing-length model \citep{panofsky1964change, ghaisas2020predictive}, or by linearly blending the corresponding upstream and downstream mean velocity profiles \citep{Chamorro2009}. The non-equilibrium behaviour of the adjusting flow also manifests in turbulence statistics, such as the higher magnitude of Reynolds normal and shear stresses and the dissipation rate of turbulent kinetic energy  \citep{antonia1972response, ismail2018simulations}. For rough-to-smooth flows, the increased magnitude of the `inner-peak' in the streamwise turbulence intensity profile located approximately 15 viscous units above the wall has been attributed to the superposition of energetic rough-wall structures (centred above the internal layer) onto the near-wall region \citep{ismail2018simulations, MogengJFM2019}.

We consider a turbulent boundary layer in the fully-rough regime as the upstream flow condition prior to the roughness transition. It can be characterised by friction Reynolds number $Re_{\tau 0}\equiv U_{\tau 0}\delta_0/\nu$ and roughness Reynolds number $k_{s0}^+\equiv U_{\tau 0}k_s/\nu$ \citep{jimenez2004turbulent}. The subscript $(\cdot)_0$ denotes flow quantities on the rough wall just prior to the transition. $U_{\tau 0}$ is the rough-wall friction velocity, $\delta_0$ is the boundary layer thickness at the rough-to-smooth change (boundary layer thickness $\delta_{99}(\hat{x})$ is defined as the wall distance where the mean streamwise velocity reaches 99\% of the freestream velocity, and $\delta_0\equiv\delta_{99}(0)$), and $\nu$ is the kinematic viscosity of air. For the rough-to-smooth change investigated here, $k_{s0}^+$ provides the magnitude of roughness change, and $Re_{\tau 0}$ reflects the outer length scale. 

Most of the existing experimental studies of a streamwise rough-to-smooth change include a small number of cases achieved by varying the freestream velocity or the morphology of the surface roughness. Both $Re_{\tau 0}$ and $k_{s0}^+$ vary concurrently as a result, and the change of roughness morphologies may also contribute to the difference observed between cases. Although classical models such as by \citet{Elliott1958} can capture the flow response with some success when the IBL occupies only a small fraction of the entire boundary layer, there are questions regarding the validity of such models in the far field where the friction velocity scale of the outer layer has been found to decay \citep{Hanson2016}. In addition, the validity of skin-friction measurements with conventional `indirect' methods such as the Clauser-chart method or Preston tube \citep{Patel1965} is compromised within a few boundary layer thicknesses downstream of the roughness transition, where such methods rely on erroneous assumptions of canonical smooth-wall behaviours beyond the EL \citep{loureiro2010distribution, MogengJFM2019}. Furthermore, the growth rate of $\delta_i$ with increasing streamwise fetch $\hat{x}$, especially the exponent $b_0$ of an assumed power law $\delta_i\propto\hat{x}^{b_0}$ has been under constant debate. For instance, \cite{Hanson2016} observed $\delta_i\propto\hat{x}^{0.36}$ for two different types of upstream roughness, \cite{mulhearn1978wind} reported $\delta_i\propto\hat{x}^{0.8}$, and through a channel flow large-eddy simulation, \cite{Saito2014} found $b_0\approx0.6$ with a slight increasing trend with increasing Reynolds number and decreasing relative roughness ($k_s/\delta$, where $\delta$ is the channel half-height). Even for the same dataset, the power-law exponent can vary depending on the method used to extract $\delta_i$ from the flow statistics \citep{rouhi2018}, further hindering the comparison of $\delta_i$ from different studies. In an attempt to redress these issues, here we design an experimental campaign which takes two cuts through the parameter space to study the effect of $Re_{\tau 0}$ and $k_{s0}^+$ independently, with a direct measure of the wall shear stress from oil film interferometry. The experiments are performed in the High Reynolds Number Boundary Layer Wind Tunnel (HRNBLWT) with a working section of 27 m, allowing measurements up to 112 boundary layer thicknesses downstream of the roughness transition where a full recovery of the energy spectrum is observed. With the aid of this dataset, we are also able to assess the blending velocity model of \cite{Chamorro2009} for a range of parameters and suggest an improvement.

The coordinate system $x$, $y$ and $z$ denotes the streamwise, spanwise and wall-normal directions. The corresponding mean velocity components are $U$, $V$ and $W$. Fluctuating velocity components are denoted by the lower case. $U_{\tau 2}$ denotes the friction velocity of the downstream surface. The subscript $(\cdot)_0$ denotes the flow quantities obtained just prior to the roughness transition. 

\section{Details of the experimental campaign} \label{sec:detail_exp}
\begin{table}
\centering

\begin{tabular}{r c c S[table-format=5.0] c c S[table-format=2.1] c c c c c l}
&Case  & Sym.  & {$Re_{\tau 0}$}  & $k^+_{s0}$ & $M$ & {$x_0$} & $U_{\infty}$ & $\delta_0 $ & $l^+$ & $\Updelta t^+$ & $\widetilde{T}_s$&\\
&         &         &                           &                                             &  & {(m)} & (m s$^{-1}$) &            (m)    &          &         & &\\
{\color{GroupRe}\ldelim\{{4}{14mm}[\emph{Group-Re}]}&\texttt{Re07ks16}     &  \TriPinkOne  & 7100     & 158  &$-3.06$&  4.5  &  21.5  & 0.11 & 22 & 0.61 & 23.7 &\\
&\texttt{Re10ks16}     &  \TriPinkTwo  & 10400   & 165   & $-3.08$&  7.2  &  22.5  & 0.15 & 24 & 0.64 & 26.4 &\\
&\texttt{Re21ks16}     &  \TriPinkFour  & 21000   & 157  & $-3.05$ & 17.1  &  24.3  & 0.32 & 24 & 0.70 & 22.7 &\\
&\texttt{Re14ks16}     &  \TriThree  & 14000   & 157   & $-3.08$ & 11.1  &  23.3 & 0.22 & 23 & 0.68 & 22.7&{\color{Groupks}\rdelim\}{3}{13mm}[\emph{Group-ks}]}\\

&\texttt{Re14ks11}     &  \CircBlueOne  & 14700   & 111  & $-2.73$ & 17.1  &  17.0  & 0.32 & 18 & 0.35 & 21.4 &\\
&\texttt{Re14ks22}     &  \DmdBlueThree  & 14500   & 228  & $-3.41$ &  7.2    &  31.0  & 0.15 & 20 & 1.26 & 24.3 &\\

\end{tabular}

\caption{Summary of the experimental cases. The friction velocity $U_{\tau 0}$ employed in calculating $Re_{\tau 0}$ and $k^+_{s0}$ is obtained over the rough fetch just upstream of the rough-to-smooth transition. The viscous-scaled hotwire sensor length $l^+$ and the sampling interval $\Updelta t^+$ are calculated using the friction velocity at the most downstream measurement location on the smooth surface. $\widetilde{T}_s \equiv T_s U_{\infty}/\delta_{99}\times 10^{-3}$, where $T_s$ is the hotwire sampling time. Note that case \texttt{Re14ks16} is shared between \emph{Group-Re} and \emph{Group-ks}, therefore its symbol can take either pink or blue colour in the corresponding group.}
\label{tab:cases}
\end{table}

\begin{figure}
\centering
\setlength{\unitlength}{1cm}
\begin{picture}(12,8)
	\put(0,0){\includegraphics{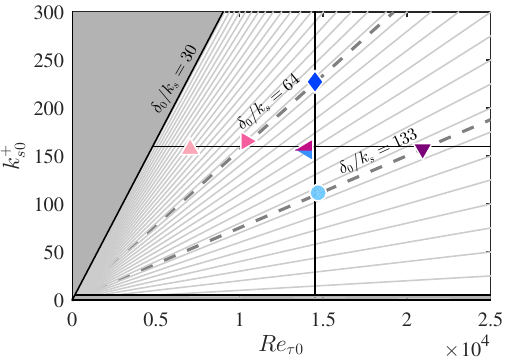}}
	\put(8.8,4.0){\emph{Group-Re}}
	\put(8.7,4.1){\vector(-1,0){0.4}}
	\put(5.7,6.6){\emph{Group-ks}}
	\put(5.6,6.7){\line(-1,0){0.3}}
	\put(5.3,6.7){\vector(0,-1){0.3}}
	\put(8.8,1.4){aerodynamically smooth}
	\put(8.7,1.5){\vector(-1,0){0.4}}
	\put(2.6,7.1){outer-layer similarity breakdown}
	\put(2.5,7.2){\line(-1,0){0.5}}
	\put(2,7.2){\vector(0,-1){1.5}}
\end{picture}
	\caption{Flow conditions ($Re_{\tau 0}$ and $k_{s0}^+$) at the immediate upstream of the roughness transition of all cases. All symbols are defined in table \ref{tab:cases}. The horizontal line is at $k_{s0}^+ = 160$, and the vertical line is at $Re_{\tau 0} = 14500$. Each light grey ray represents a combination of constant $\delta_0/k_s$, the reciprocal of which decreases from 0.033 ($\delta_0/k_s = 130$) to 0.001 ($\delta_0/k_s = 1000$) with a constant step of 0.001 in the clockwise direction. The two dashed lines show the cases with matched $\delta_{0}/k_s$ of 64 and 133.}
	\label{fig:CaseSummary}
\end{figure}
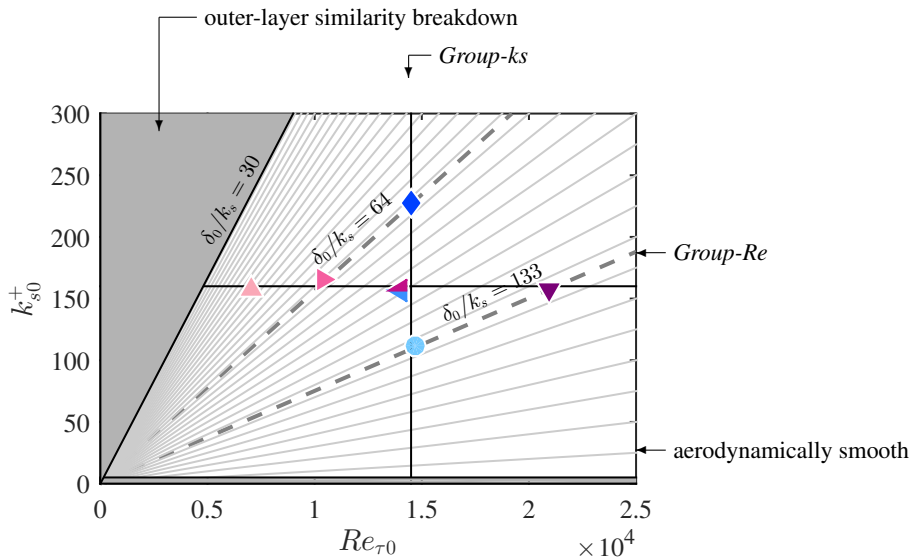

Two groups of wind tunnel experiments are designed to examine the effect of $Re_{\tau 0}$ and $k_{s0}^+$ on the flow recovery behaviour separately. \emph{Group-Re} consists of measurements with varying $Re_{\tau0}$ and $k_{s0}^+$ held constant, while \emph{Group-ks} measurements vary $k_{s0}^+$ while holding $Re_{\tau0}$ constant. The same type of sandpaper (P24 grit) is used in all cases, which ensures a constant $k_s$, while $x_0$, the downstream location of the roughness transition is varied. For \emph{Group-Re} measurements, the freestream velocity $U_{\infty}$ is adjusted to account for the gradual decrease of the skin-friction coefficient $C_f(\equiv \tau_w/\frac{1}{2}\rho U_{\infty}^2)$ with Reynolds number, to maintain an approximately constant $U_{\tau 0}$ at the rough surface immediately upstream of the rough-to-smooth transition. This will guarantee a nominally constant $k_{s0}^+$ for all cases. The variation of $Re_{\tau0}$ is primarily achieved by varying the $x_0$ location of the transition. For \emph{Group-ks} measurements, $U_{\infty}$ is adjusted to account for the growth of $\delta_{99}$ with $x_0$ and maintain a constant $Re_{\tau0}$. This will subsequently lead to a variation in $k_{s0}^+$. The aforementioned variation of flow parameters in each group is summarised in figure \ref{fig:CaseSummary} and the relevant experimental conditions are listed in table \ref{tab:cases}. The streamwise length of the sandpaper, $x_0$, is more than 40 times greater than $\delta_0$ in the shortest case (\texttt{Re07ks16}), which is greater than the rough fetch used in many well-accepted studies \citep[][for instance]{antonia1972response, Hanson2016}, and the rough-to-smooth transition is sufficiently downstream of any expected tripping effects or inlet artefacts \citep{marusic2015evolution}. It is also worth mentioning that $\delta_0/k_s$ is greater than 40 in all cases (see figure \ref{fig:CaseSummary}) and so the canonical outer-layer similarity is expected for the oncoming rough-wall boundary layers.  Each case is assigned a code in the format of \texttt{Re\textit{xx}ks\textit{yy}}, where \texttt{\textit{xx}}$\approx Re_{\tau 0}/1000$, and \texttt{\textit{yy}}$\approx k^+_{s0}/10$. \emph{Group-Re} consists of cases \texttt{Re07ks16}, \texttt{Re10ks16}, \texttt{Re14ks16} and \texttt{Re21ks16}, in which a nominally constant $k_{s0}^+\approx 160$ is maintained while $Re_{\tau 0}$ increases from 7100 to 21000. \emph{Group-ks} consists of cases \texttt{Re14ks11}, \texttt{Re14ks16} and \texttt{Re14ks22}, in which a nominally constant $Re_{\tau 0} \approx 14000$ is maintained while $k_{s0}^+$ increases from 111 to 228. As shown in table \ref{tab:cases}, a unique symbol is assigned to each case, which will be adhered to throughout this paper (unless there are exceptions). All cases in \emph{Group-Re} are represented by symbols with different shades of magenta, while cases in \emph{Group-ks} are represented by symbols with different shades of blue.  The shading of symbols darkens with increasing $Re_{\tau 0}$ in \emph{Group-Re} and $k_{s0}^+$ in \emph{Group-ks}. Note that case \texttt{Re14ks16} is at the `crossing point' in the parameter space so it is shared between two groups, therefore its symbol can take either pink or blue when plotted as part of the corresponding group.

\begin{figure}
    \centering
    \includegraphics[scale = 0.72, trim = 8mm 0mm 0 10mm,clip]{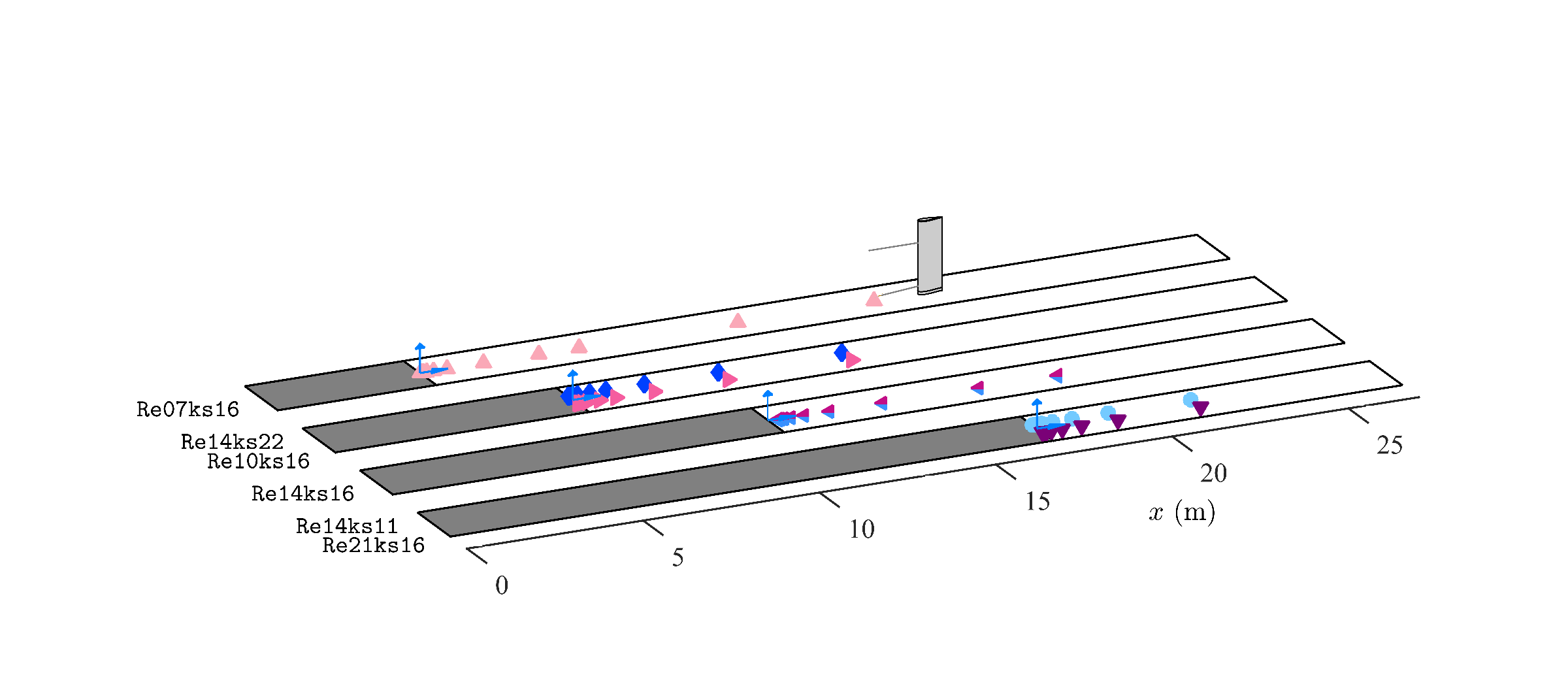}
    \put(-195,115){\footnotesize{\ding{172}}}  \put(-188,106){\footnotesize{\ding{173}}}
    \put(-330,120){\footnotesize{\ding{172}: Pitot tube }}
    \put(-330,110){\footnotesize{\ding{173}: hotwire probe}}
    \caption{Overview of the experimental setup. The flow is going from the left to the right of the figure. The grey shaded surface represents the sandpaper, and the white region represents the smooth wall. Streamwise locations where a wall-normal hotwire profile is acquired in each case are shown by the corresponding symbols.}
    \label{fig:ExpSetup}
\end{figure}
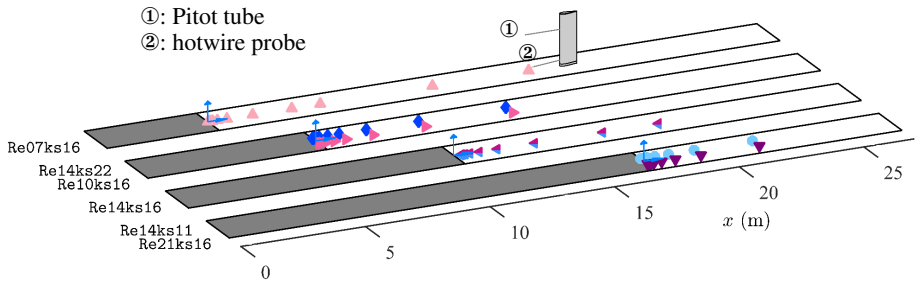

\subsection{The facility}
\begin{figure}
    \centering
\includegraphics[width=0.9\textwidth, trim = 0mm 0mm 0 0mm,clip]{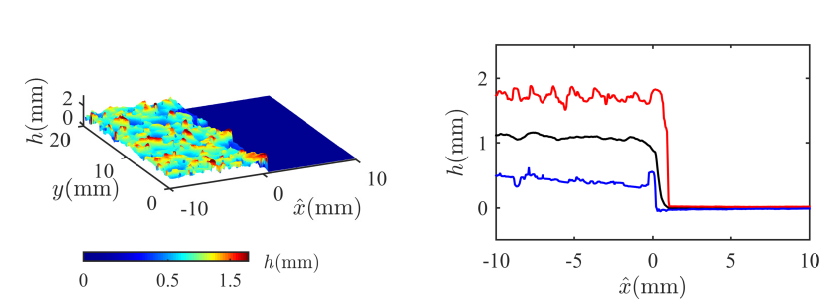}
\put(-360,100){(\textit{a})}
\put(-165,100){(\textit{b})}
\put(-110,72){average}
\put(-110,92){{\color{red}max}}
\put(-110,54){{\color{blue}min}}
    \caption{(\textit{a}) The surface elevation at the rough-to-smooth transition measured using an in-house built laser scanner. The black line in (\textit{b}) is the spanwise average of the surface elevation, and the red and blue lines are the maximum and minimum of the surface elevation along each spanwise line.}
    \label{fig:P24roughness}
\end{figure}
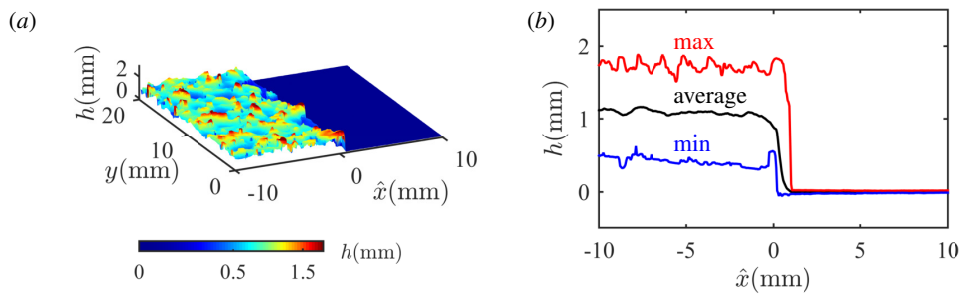

The experiments are performed in the High Reynolds Number Boundary Layer Wind Tunnel (HRNBLWT) at the University of Melbourne, which has a $27\times1.89\times0.92$ m$^3$ working section (length $\times$ width $\times$ height), see \cite{Kulandaivelu2012} and \cite{marusic2015evolution} for further details. The overview of the experimental setup is depicted in figure \ref{fig:ExpSetup}. An upstream portion of the tunnel floor in the test section is covered by P24 grit sandpaper (SP40F, Awuko Abrasives) from the inlet to the location of $x_0$ (as shown by the grey coloured patches in figure \ref{fig:ExpSetup}), while the remaining length is the original smooth aluminium surface. The sandpaper patch has a width of 1.82 m, covering nearly the entire width of the tunnel. By adjusting the bleeding slots on the tunnel roof, a nominal zero pressure gradient condition ($C_p(x)\equiv1-\left(U_{\infty}(x)/U_{\mathrm{ref}}\right)^2=0\pm0.01$) is achieved, where $U_{\mathrm{ref}}$ is obtained from an NPL (National Physical Laboratory) Pitot-static tube fixed in the freestream at $x = 0.7$ m from the tunnel inlet, and $U_{\infty}(x)$ from another Pitot-static tube mounted on the sting and traversed along the test section. To characterise the roughness parameters, a $60 \,\mathrm{mm} \times 60 \,\mathrm{mm}$ patch of the rough-to-smooth transition is scanned using an in-house built laser scanner, which consists of a Keyence LK-031 laser triangulation sensor with a spot diameter of 0.03 mm and a linearity of 0.01 mm attached to an $x-y$ traverse with a step size of 0.05 mm in both directions. The resulting surface topography is shown in figure \ref{fig:P24roughness}. The smooth wall is approximately 1.8 mm below the roughness crest, i.e. the step height between the roughness crest and the smooth surface downstream is $\Updelta H = -1.8 \,\mathrm{mm}$. Detailed roughness parameters of the P24 sandpaper are listed in table \ref{tab:roughness}.

\begin{table}
\centering
\begin{tabular}{ c c c c}
Roughness parameter & Value  & Units & Formula\\
$k$					& 1.424 & mm   & $6\sqrt{\overline{h'^2}}$\\
$k_a$                        & 0.182 & mm  & $\overline{|h'|}$\\
$k_{z,5\times5}$		& 1.614 & mm   & $\overline{\max{h'_{5\times5}}-\min{h'_{5\times5}}}$\\
$k_{rms}$			& 0.237 & mm   & $\sqrt{\overline{h'^2}}$\\
$k_{sk}$			& 0.222 & -        & $\overline{h'^3}/k_{rms}^3$\\
$k_{ku}$			& 3.912 & -        & $\overline{h'^4}/k_{rms}^4$\\
$ES_x$				& 0.610 & - 	     & $\overline{\left|\frac{\mathrm{d}h'}{\mathrm{d}x}\right|}$\\

\end{tabular}
\caption{Roughness parameters from the scanned surface elevation of the P24 sandpaper. Here $h'\equiv h-\overline{h}$ is the surface deviation about the mean height, and $\max{h'_{5\times5}}$ and $\min{h'_{5\times5}}$ are the maximum and minimum of $h'$ in each of the $5\times5$ tiles of the scanned surface \citep{thakkar2017surface}.}
\label{tab:roughness}
\end{table}

\subsection{Hotwire anemometry}

A conventional single-wire hotwire probe is operated by an in-house designed Melbourne University Constant Temperature Anemometer (MUCTA). To minimise aliasing, the hotwire signal is filtered using a Frequency Devices 9002 analogue filter with a cut-off  frequency set to half of the sampling frequency. Digitisation of the resulting signal is via a Data Translation DT9836 data acquisition board with sampling frequency 50 kHz. The sampling parameters are summarised in table \ref{tab:cases}. The viscous-scaled sampling interval $\Updelta t^+$ is less than 2 for all cases, which is safely below the threshold of $\Updelta t^+\approx 3$ to ensure a temporally well resolved signal \citep{Hutchins2009}. The hotwire sampling time $T_s$ is more than 20000 boundary layer turn-over time ($\delta_{99}/U_{\infty}$) to achieve a good convergence of the statistics. Calibration is performed following an \textit{in-situ} procedure before and after each measurement. Thereafter, any drift is corrected by an intermediate single point re-calibration (ISPR) method discussed in \citet{talluru2014calibration}, where the hotwire voltage is periodically monitored in the freestream. The uncertainty in $U$ and $\overline{u^2}$ is usually within $1\%$ and $3\%$, respectively \citep{yavuzkurt1984guide}. The method of calibration drift correction proposed by \citet{talluru2014calibration} employed here offers further improvements. During each run, the air temperature and atmospheric pressure data are also sampled to calculate $\nu$, the kinematic viscosity of air \citep{sutherland1893lii}. The resulting $\nu$ varies from $1.51\times10^{-5}$ m$^2$/s to $1.64\times10^{-5}$ m$^2$/s in different runs.

\begin{figure}
    \centering
\setlength{\unitlength}{1cm}
\begin{picture}(13,5.7)
\put(0,0){\includegraphics{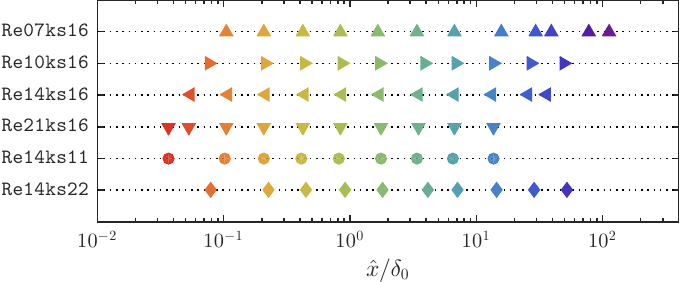}}
\end{picture}
    \caption{Downstream fetches where hotwire measurements are performed for all cases. The shape of each symbol indicates the case following the convention of table \ref{tab:cases}, while the colour reflects the magnitude of $\hat{x}/\delta_0$.}
    \label{fig:x_d_color}
\end{figure}

Velocity profiles are obtained by traversing a single-normal hotwire probe over 40 logarithmically spaced wall-normal locations from $z/\delta_{99} \approx 0.001$ to 2. The sensing element of this probe has a diameter of $d = 2.5\,\upmu$m and a length-to-diameter ratio of 200. Note that in case \texttt{Re14ks22} where $U_{\tau 0}$ is the highest among all cases, a wire with a smaller diameter ($d = 1.5\,\upmu$m) is used to maintain a similar spatial resolution in wall units \citep{Hutchins2009}. Wall-normal boundary layer profiles are measured at over 10 logarithmically spaced streamwise locations downstream of the rough-to-smooth transition, from $\hat{x} = 12\,\mathrm{mm}$ to $x = 21\,\mathrm{m}$, as shown in figure \ref{fig:ExpSetup} and also on a logarithmic axis in figure \ref{fig:x_d_color}. The colour of each symbol in figure \ref{fig:x_d_color} is determined by its corresponding $\hat{x}/\delta_0$. This colour scheme will be used in the study of the streamwise evolution of the flow where a series of profiles at various streamwise locations are shown in one figure. A reference profile above the rough surface is also acquired at $\hat{x} = -0.1\,\mathrm{m}$ in each case. In order to accurately measure the wall location, a Renishaw RGH24 optical linear encoder with a resolution of 1 $\upmu$m is used in the HRNBLWT. The initial offset is determined using a wall-normal traversing microscope equipped with a digital micrometer (CDI BG3600) with a resolution of 1 $\upmu$m.

\subsection{Oil-film interferometry} \label{sec:OFI}
The wall-shear stress $\tau_w$ is directly measured using Oil-Film Interferometry  (OFI), which is one of the few methods available for a direct measurement at the surface with the required streamwise resolution to capture the rapid evolution following the rough-to-smooth transition. The technique measures the thinning rate of an oil film as it is being acted upon by the shear near the wall, which in turn permits an accurate measure of the mean wall-shear stress \citep{Tanner1976, fernholz1996new, zanoun2003evaluating}. The thickness of a typical oil film is in the order of micrometres, which can be measured by the fringe pattern from the interference of light reflected from the top and bottom of the oil film. Moreover, the technique utilises inexpensive equipment, which includes a consumer camera and monochromatic light source (or a non-monochromatic light source with a bandpass filter).
 
The experimental procedure in this study follows a similar process as described in \cite{MogengJFM2019}. A 1.4 m long and 1 m wide glass insert has been installed in the tunnel floor at $x = 5\,\mathrm{m}$, providing optical access from the underside of the tunnel for $\hat{x}/\delta_0 < 7 $ in case \texttt{Re07ks16}. Both glass and aluminium surfaces are aerodynamically smooth. Steps between the glass insert and the aluminium wall are limited to 0.05 mm (2 wall units). The upstream joins between the glass and the aluminium floor are beneath the sandpaper. The OFI measurements are conducted on the centreline of the tunnel floor, therefore they are at a sufficient distance ($>5\delta_0$) away from these joins. This configuration provides a well-resolved fringe pattern with approximately 50 pixels per wavelength. A line of silicone oil is placed along the spanwise direction on the glass surface and spread downstream by the wind shear. The oil film is illuminated by an Imalent DX80 LED torch, and recorded using a Nikon D810 DSLR camera with a Tamron $180\,\mathrm{mm}$ macro lens. A $532\,\mathrm{nm}$ bandpass filter with a bandwidth of 10 nm is attached to the camera lens to obtain monochromatic fringe patterns. For the remaining measurements, a glass insert on the centerline of the working section ceiling provides optical access from above. To improve the fringe quality, the tunnel floor is covered by a piece of black mylar film with a thickness less than $40\, \upmu$m (equivalent to 1.5-2.5 wall units for all cases). This small step in the surface is at least 2 cm ($>500$ mylar film thicknesses, $>750$ wall units) away from the oil droplet in all directions except for the first few downstream locations, where the mylar film extends all the way to the rough-to-smooth change. The same illumination and imaging system as in the previous configuration is used, but with a reduced resolution of approximately 30 pixels per wavelength due to the $1\, \mathrm{m}$ stand-off distance between the camera and oil film. Both OFI configurations (imaging from underneath and above) have been compared at $\hat{x}/\delta_0 = 4$ for case \texttt{Re07ks16}, and are shown to give the same result to within $1\%$.

For both configurations, 100 images are captured with a time interval of five seconds in each measurement. The camera calibration and image processing algorithm are the same as detailed in \cite{deSilva2018}.


\begin{figure}
\centering
\setlength{\unitlength}{1cm}
\begin{picture}(7,5.3)
\put(0,0){\includegraphics[scale = 1]{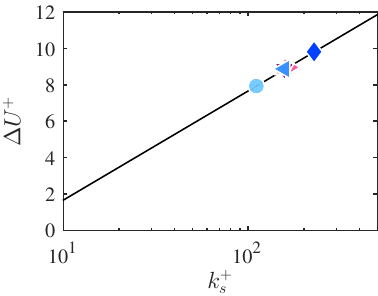}}
\end{picture}
\caption{Hama function, $\Updelta U^+$, as a function of the inner-scaled equivalent sand grain roughness $k_s^+$ for all cases. All symbols are defined in table \ref{tab:cases} with the colour indicating $Re_{\tau 0}$ or $k_{s0}^+$, and the solid line is Nikuradse's fully-rough relationship as shown in (\ref{eq:Nikuradse}).}
\label{fig:Rough_ksp}
\end{figure}
\section{Initial flow conditions on the rough surface} \label{sec:Rough}

The streamwise evolution of the turbulent boundary layer starts from the initial flow condition on the rough surface. Once a rough-wall profile is specified, the flow recovery downstream on a smooth surface can be fully determined. A wall-normal hotwire profile is obtained at $\hat{x} = -0.1$ m in each case to document the initial condition. For the hotwire measurements on the rough surface, $z = 0$ is located at the roughness crest following the method of measuring the initial offset as detailed in \cite{squire2016comparison}. Following the same study, the virtual origin of the roughness is assumed to be at the averaged surface elevation $\overline{h}=1.2$ mm (figure \ref{fig:P24roughness}\textit{b}), therefore a wall positioning correction of $\varepsilon =\Updelta H+\overline{h} =  -0.6$ mm is used. In addition, \citet{squire2016comparison} also demonstrated that for  a similar type of sandpaper, varying the virtual origin from the roughness crest to the trough leads to a less than $0.8\%$ change in $U_{\tau}$ downstream of $x = 4.75$ m.

Nikuradse's (\citeyear{nikuradse1950laws}) equivalent sand grain roughness $k_s$ of the surface is obtained from the highest Reynolds number case \texttt{Re21ks16} following the procedures below: firstly, $U_{\tau 0}$ on the rough surface is estimated by enforcing an outer-layer similarity \citep{Townsend1976} in the mean velocity deficit profile for $z/\delta_{99}>0.3$. The Hama function, $\Updelta U^+$, which is the vertical shift between the rough-wall profile and the logarithmic law $U^+ = \frac{1}{\kappa}\ln(z^+)+B$, is then determined by minimising the least-squares error in the inertial range with the upper and lower bounds given in \cite{mehdi2013mean} and \cite{squire2016comparison}. The constants in the logarithmic law are chosen as $\kappa = 0.384$ and $B = 4.17$ \citep{Nagib2007}. Finally, $k_{s0}^+$ is computed from $\Updelta U^+$ assuming Nikuradse's fully-rough relationship
\begin{equation}
\Updelta U^+ = \frac{1}{\kappa}\ln(k_{s0}^+)+B-A'_{FR},
\label{eq:Nikuradse}
\end{equation}
where $A'_{FR} = 8.5$ \citep{nikuradse1950laws}. The equivalent sand grain roughness is found to be $k_s \approx 2.43$ mm, which is approximately $1.7k$. 

This $k_s$ value is assumed for all other cases obtained with the same sandpaper. The rough-wall friction velocity $U_{\tau 0}$ is determined by minimising the difference between the Hama function obtained from the inner-normalised mean velocity profile and from the fully-rough relationship (\ref{eq:Nikuradse}). The applicability of (\ref{eq:Nikuradse}) can be justified in this case since all the rough-wall measurements are planned in the fully-rough regime with $k_{s0}^+>100$. Note that this method to determine $U_{\tau0}$ from the measured mean velocity profile is similar to the approach described by \cite{squire2016comparison}. It is essentially a more informed modified Clauser chart method with $k_s$ (and therefore the relationship between $\Updelta U^+$ and $U_{\tau0}$) prescribed. $\Updelta U^+$ and $k_{s0}^+$ for all cases determined by this approach are shown in figure \ref{fig:Rough_ksp}. In \emph{Group-ks} where the effect of $k_{s0}^+$ is studied, $k_{s}^+$ varies from 111 to 228, corresponding to a range of $\Updelta U^+$ from 7.9 to 9.8. 

\begin{figure}
\centering
\setlength{\unitlength}{1cm}
\begin{picture}(13,5.1)
\put(0,0){\includegraphics[scale = 0.95]{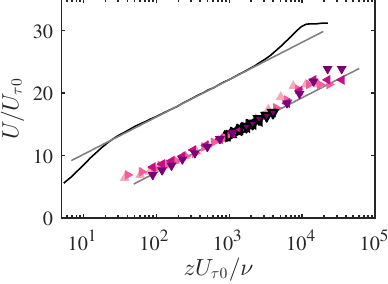}}
\put(6.5,0){\includegraphics[scale = 0.95]{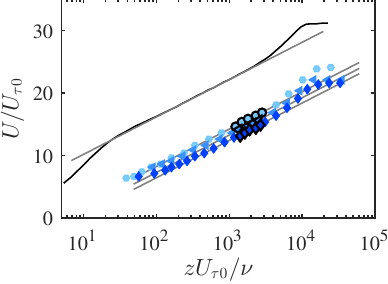}}
\put(0.1,4.8){(\textit{a})}
\put(6.6,4.8){(\textit{b})}
\end{picture}

\caption{Inner-normalised mean streamwise velocity profiles at $\hat{x} = -0.1$ m from (\textit{a}) \emph{Group-Re} and (\textit{b}) \emph{Group-ks}. The black line in both figures is a smooth-wall reference profile with $Re_{\tau} = 1.0\times10^4$ acquired in the same facility using hotwire anemometry and normalised using the corresponding smooth-wall friction velocity \citep{marusic2015evolution}. The colour of the symbols indicates (\textit{a}) $Re_{\tau 0}$ and  (\textit{b}) $k_{s0}^+$, as defined in table \ref{tab:cases}. Symbols with a black outline represent the extent of the inertial range. These data points are employed in the fit to obtain $U_{\tau 0}$. Only every second data point is shown for clarity except in the inertial range.}
\label{fig:Rough_U_inner}
\end{figure}

\begin{figure}
\centering
\setlength{\unitlength}{1cm}
\begin{picture}(13,5.1)
\put(0,0){\includegraphics[scale = 0.95]{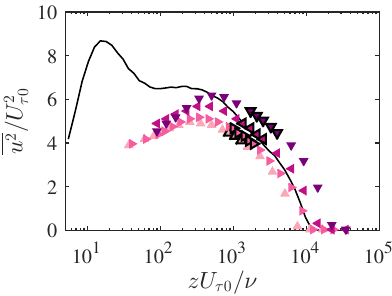}}
\put(6.5,0){\includegraphics[scale = 0.95]{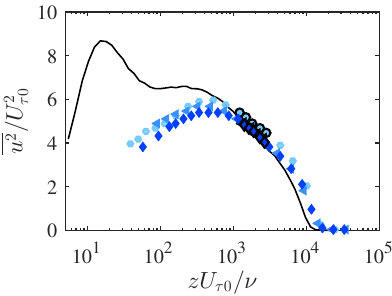}}
\put(0.1,4.8){(\textit{a})}
\put(6.6,4.8){(\textit{b})}
\end{picture}

\caption{Inner-normalised streamwise turbulence intensity profiles at $\hat{x} = -0.1$ m from (\textit{a}) \emph{Group-Re} and (\textit{b}) \emph{Group-ks}. Legends are the same as in figure \ref{fig:Rough_U_inner}.}
\label{fig:Rough_uvar_inner}
\end{figure}

\begin{figure}
\centering
\setlength{\unitlength}{1cm}
\begin{picture}(13,10.2)
\put(0,0){\includegraphics[scale = 0.95]{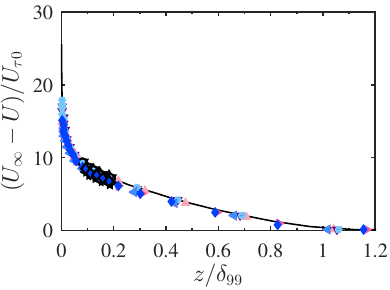}}
\put(6.5,0){\includegraphics[scale = 0.95]{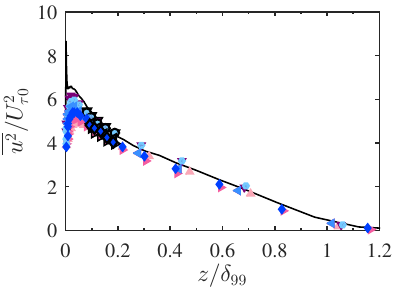}}
\put(0,5.1){\includegraphics[scale = 0.95]{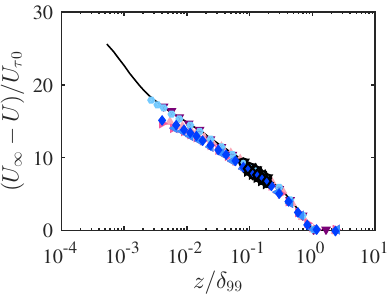}}
\put(6.5,5.1){\includegraphics[scale = 0.95]{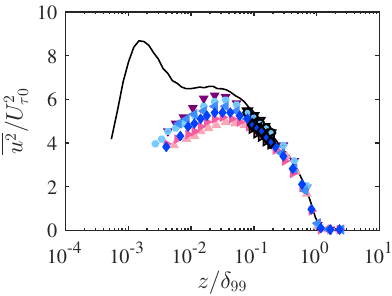}}
\put(0.1,9.9){(\textit{a})}
\put(6.6,9.9){(\textit{b})}
\put(0.1,4.8){(\textit{c})}
\put(6.6,4.8){(\textit{d})}
\end{picture}
\caption{(\textit{a}) Mean streamwise velocity deficit and (\textit{b}) streamwise turbulence intensity versus outer-normalised wall-normal distance for both \emph{Group-Re} and \emph{Group-ks}. The corresponding plots with a linear $x$-axis are shown as (\textit{c}) and (\textit{d}). Legends are the same as in figure \ref{fig:Rough_U_inner}. }
\label{fig:Rough_OutSim}
\end{figure}
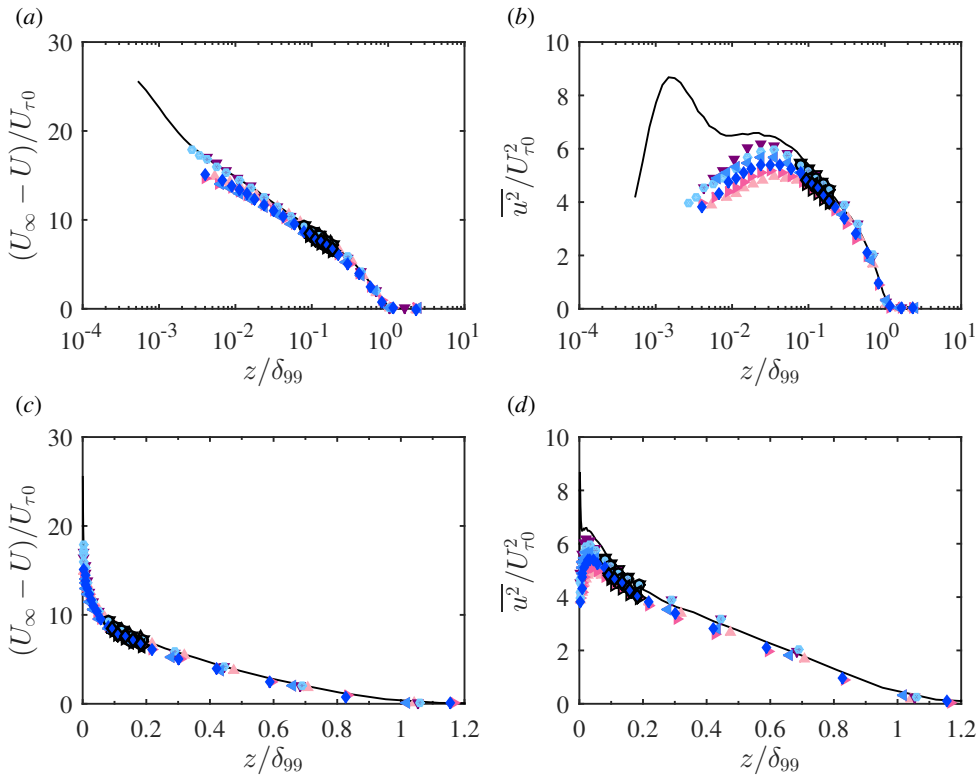

Figures \ref{fig:Rough_U_inner} and \ref{fig:Rough_uvar_inner} respectively show the inner-normalised mean streamwise velocity and turbulence intensity profiles of both \emph{Group-Re} (\textit{a}) and \emph{Group-ks} (\textit{b}) just upstream of the roughness transition at $\hat{x} = -0.1$ m. As expected, a good collapse of the data in the logarithmic region in figure \ref{fig:Rough_U_inner}(\textit{a}) indicates that $\Updelta U^+$ (and therefore $k_{s0}^+$) in \emph{Group-Re} are closely matched, whereas the vertical shifts in the velocity profile in figure \ref{fig:Rough_U_inner}(\textit{b}) correspond to the $k_{s0}^+$ trend as planned for \emph{Group-ks}. Similarly the $Re_{\tau 0}$ trend can be observed in figure \ref{fig:Rough_U_inner}(\textit{a}) and figure \ref{fig:Rough_uvar_inner}(\textit{a}). Townsend's outer-layer similarity hypothesis \citep{Townsend1976} appears to be nominally satisfied as shown by the outer-normalised profiles for both groups in figure \ref{fig:Rough_OutSim}, in agreement with previous studies of three-dimensional roughness \citep[for example][]{flack2005experimental, wu2007outer, squire2016comparison}. Overall, these results demonstrate that the rough-wall profiles immediately upstream of the roughness transition exhibit similar behaviours as a turbulent boundary layer developed on a homogeneously rough surface in a quasi-equilibrium state.

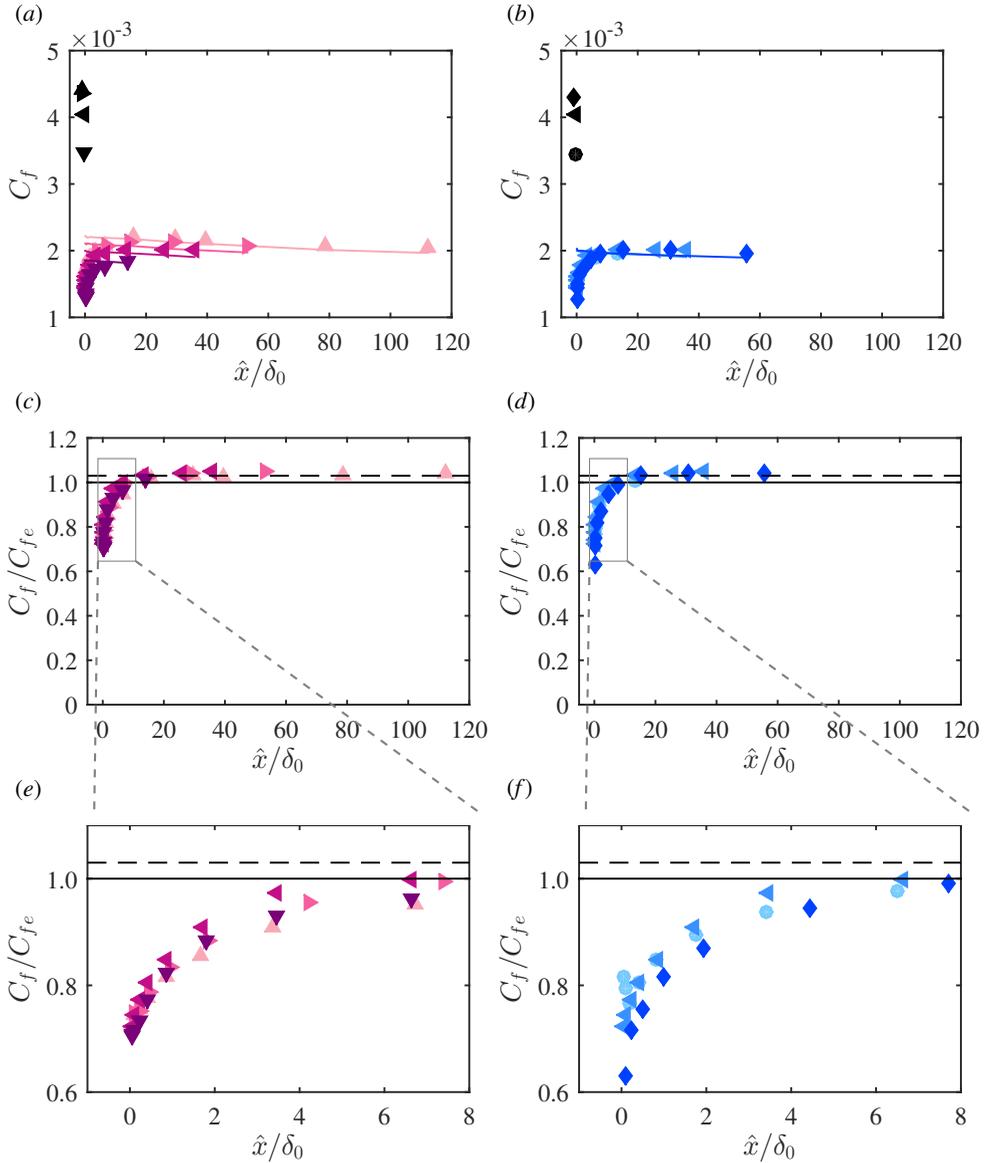
\begin{figure}
	\centering
	\setlength{\unitlength}{1cm}
	\begin{tikzpicture}(13,15.3)
	\node[anchor=south west,inner sep=0] at (0,10.2){\includegraphics[scale = 0.95]{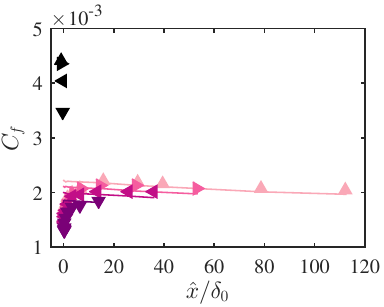}};
	\node[anchor=south west,inner sep=0] at (6.5,10.2){\includegraphics[scale = 0.95]{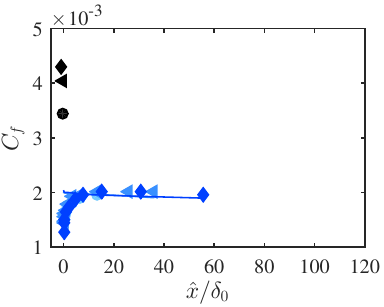}};
	\node[anchor=south west,inner sep=0] at (0,5.1){\includegraphics[scale = 0.95]{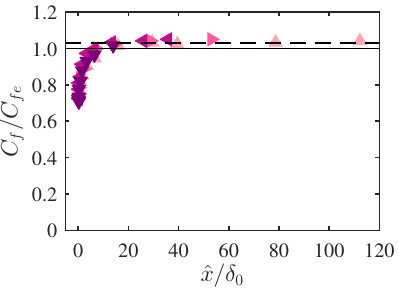}};
	\node[anchor=south west,inner sep=0] at (6.5,5.1){\includegraphics[scale = 0.95]{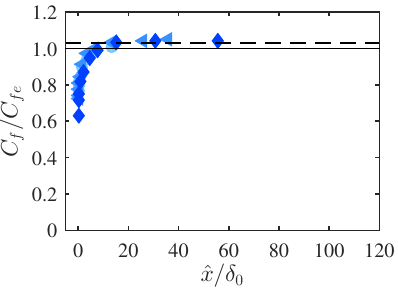}};
	\node[anchor=south west,inner sep=0] at (0,0){\includegraphics[scale = 0.95]{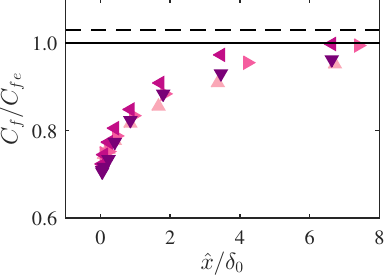}};
	\node[anchor=south west,inner sep=0] at (6.5,0){\includegraphics[scale = 0.95]{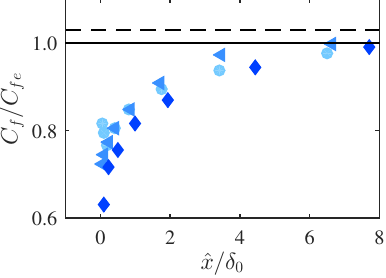}};

	\node[anchor=south west,inner sep=0] at (0.1,15){(\textit{a})};
	\node[anchor=south west,inner sep=0] at (6.6,15){(\textit{b})};
	\node[anchor=south west,inner sep=0] at (0.1,9.9){(\textit{c})};
	\node[anchor=south west,inner sep=0] at (6.6,9.9){(\textit{d})};
	\node[anchor=south west,inner sep=0] at (0.1,4.8){(\textit{e})};
	\node[anchor=south west,inner sep=0] at (6.6,4.8){(\textit{f})};

	\draw[gray] (1.2,7.95) -- (1.7,7.95) -- (1.7,9.3) -- (1.2,9.3) -- (1.2,7.95);
	\draw[gray] (7.7,7.95) -- (8.2,7.95) -- (8.2,9.3) -- (7.7,9.3) -- (7.7,7.95);
	\draw[gray, thick, dashed] (1.2,7.95) -- (1.15,4.6);
	\draw[gray, thick, dashed] (1.7,7.95) -- (6.3,4.6);
	\draw[gray, thick, dashed] (7.7,7.95) -- (7.65,4.6);
	\draw[gray, thick, dashed] (8.2,7.95) -- (12.8,4.6);

	\end{tikzpicture}

    \caption{Skin-friction coefficient $C_f$ versus the fetch $\hat{x}$ over the downstream smooth surface scaled by $\delta_0$ for (\textit{a}) \emph{Group-Re} and (\textit{b}) \emph{Group-ks}. The coloured symbols represent OFI measurements on the smooth wall,  and the black symbols are obtained from the reference profile over the rough surface. The solid lines show $C_{fe}$, the expected equilibrium smooth-wall skin-friction coefficient at every streamwise location. (\textit{c}) and (\textit{d}) are $C_f$ normalised by its equilibrium value $C_{fe}$ for \emph{Group-Re} and \emph{Group-ks}, respectively, with the corresponding magnified view of the boxed region shown in (\textit{e}) and (\textit{f}). The colour of the symbols and solid lines indicates $Re_{\tau 0}$ and $k_{s0}^+$ in the left and right column, respectively, as defined in table \ref{tab:cases}. The solid horizontal line is $C_f/C_{fe} = 1$, and the dashed line is $C_f/C_{fe} = 1.03$. }
    \label{fig:Cf}
\end{figure} 
\section{Skin-friction coefficient}\label{sec:big_Cf}
We first examine the recovery of the skin-friction coefficient downstream of the rough-to-smooth change for all cases. Figure \ref{fig:Cf}(\textit{a},\textit{b}) show the evolution of $C_f$ following the rough-to-smooth transition of \emph{Group-Re} and \emph{Group-ks} cases, respectively. Over the smooth surface, $C_f$ is measured directly using OFI (shown by the coloured symbols in figure \ref{fig:Cf}\textit{a},\textit{b}), and the reference $C_f$ over the rough surface (black symbols) is calculated from the mean velocity profile following the method detailed in \S\ref{sec:Rough}. In all cases, $C_f$ undershoots the expected equilibrium smooth-wall value (shown by the solid lines) immediately downstream of the roughness transition, then reaches its maximum at $\hat{x}/\delta_0\approx 20$ before decreasing gradually farther downstream as dictated by the increasing Reynolds number of the flow. The black symbols show $C_f$ on the rough surface just upstream of the surface transition. In all cases there is a three- to five-fold decrease in $C_f$ immediately downstream of the transition. When comparing $C_f$ between cases in \emph{Group-Re}, we notice that $C_f$ on both rough and smooth surfaces decreases with $Re_{\tau 0}$ as evident in figure \ref{fig:Cf}(\textit{a}). Such a dependence on $Re_{\tau 0}$ is expected and is similar to that observed for a turbulent boundary layer on a homogeneous surface \cite[e.g.][]{Nagib2007}. For \emph{Group-ks} measurements on the other hand, $C_f$ increases with $k_{s0}^+$ only on the rough surface (as shown by the solid black symbols in figure \ref{fig:Cf}\textit{b}). The smooth-wall $C_f$ from all 3 cases collapse to a single trend, as these cases have very similar $Re_{\tau}$ values on the smooth wall, and the $C_f$ of the fully recovered flow is solely determined by the local $Re_{\tau}$.

The expected equilibrium skin-friction coefficient, $C_{fe}$, is defined as the skin-friction coefficient that an equilibrium turbulent boundary layer at the same Reynolds number (based on momentum thickness) would have \citep{Hanson2016}. $C_{fe}$ is estimated using an empirical relationship obtained from drag balance measurements of a smooth-wall turbulent boundary layer in the same facility \citep{baars2016wall}:
\begin{equation} 
C_{fe} = 2\left(\ln \left(Re_{\theta}\right)/0.38+3.7\right)^{-2},
\label{Eq:Cf_eql}
\end{equation}
where $Re_{\theta}\equiv U_{\infty} \theta/\nu$, and $\theta$ is the momentum thickness computed locally by integrating the measured mean velocity profile. If the flow has fully recovered to the smooth-wall condition, then $C_{f}$ should equal to $C_{fe}$. Therefore, it can serve as an indication of the flow recovery. Similar to $C_f$ at a large fetch, $C_{fe}$ also decreases with increasing $Re_{\tau 0}$ (therefore increasing $Re_{\theta}$) in \emph{Group-Re}. As a consequence of the matched $Re_{\tau0}$, \emph{Group-ks} cases have similar $Re_{\theta}$ downstream of the rough-to-smooth change, which leads to the agreement in $C_{fe}$ between cases.

The recovery of $C_f$ towards $C_{fe}$ is further examined through the ratio $C_f/C_{fe}$ in figure \ref{fig:Cf}(\textit{c},\textit{d}). Immediately downstream of the rough-to-smooth transition, $C_f$ is approximately 60\%---80\% of $C_{fe}$ for all cases, followed by a quick recovery to $C_{fe}$ within approximately $20\delta_0$. The data points overshoot $C_{f}/C_{fe} = 1$ (the black horizontal line) slightly and then reach a plateau at $C_{f}/C_{fe} = 1.03$. This 3\% difference is possibly related to the uncertainty in the data and the empirical relationship (\ref{Eq:Cf_eql}) employed. Regardless, there seems to be little difference between cases in both \emph{Group-Re} and \emph{Group-ks} in terms of the $C_f$ recovery behaviour when scaled by $\delta_0$ and $C_{fe}$ in the far field. When normalised by $C_{fe}$ as shown in figures \ref{fig:Cf}(\textit{c},\textit{d}), to within the experimental uncertainty, $C_f$ evolves as if the flow were in quasi-equilibrium with the smooth wall beyond $\hat{x}/\delta_0\approx 20$. Utilising two different types of roughness (grit and mesh),  \cite{Hanson2016} achieved a more than five-fold change in $k_{s0}^+$, and they reported a decrease in the recovering $C_f$ with increasing $k_{s0}^+$. This range of $k_{s0}^+$ is unfortunately not attainable without physically increasing the equivalent sand-grain roughness $k_s$, and we do not observe a trend with $k_{s0}^+$ in $C_f$ as a result of the limited $k_{s0}^+$ range in this study. For a larger range of $k_{s0}^+$ (larger perturbation strength $|M|$) we would expect to see an influence on $C_f$ recovery. However, it is interesting to note that  in both studies the full recovery of $C_f$ to the smooth-wall value is found at a streamwise fetch of $\hat{x}/\delta_0\approx 20$.

The scatter in $C_f$ close to the roughness transition between cases appears to be more prominent in \emph{Group-ks} compared to \emph{Group-Re} as revealed by the magnified view in figures \ref{fig:Cf}(\textit{e},\textit{f}). As the step height between the roughness crest and the smooth surface $\Updelta H = -1.8 \,\mathrm{mm}$ is the same for all cases in both groups, this trend in $C_f$ might be dependent on the inner-normalised step height $\Updelta H^+$, which increases with $k_{s0}^+$ (effectively $U_{\tau 0}/\nu$) in the current implementation. It has also been shown by OFI measurements using a spanwise line of silicone oil at a comparable Reynolds number that the local distribution of roughness elements can have an effect on $C_f$ in the range of $\hat{x}/\delta_0<0.4$ \citep{MogengAFMC2018}. This might also be responsible for some of the variation in $C_f$ in the extreme near field ($0 < \hat{x}/\delta_0 < 0.4$) as the local roughness geometry (at an element scale) on the centreline of the tunnel floor (where the OFI measurements are performed) varies between cases where a different piece of sandpaper is used.

\section{Internal boundary layer}\label{sec:big_IBL}

\begin{figure}
    \centering
\setlength{\unitlength}{1cm}
\begin{picture}(6,7.5)
    \put(0,0){\includegraphics[scale = 0.95]{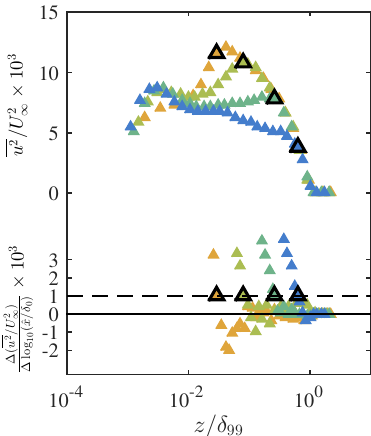}}
    \put(1.5,6){(\textit{a})}
    \put(1.5,3){(\textit{b})}
\end{picture}
    \caption{(\textit{a}) Streamwise turbulence intensity of case \texttt{Re07ks16} normalised by $U_{\infty}$, plotted against the outer-scaled wall-normal location $z/\delta_{99}$. (\textit{b}) The difference between two consecutive (in $x$)  $\overline{u^2}/U_{\infty}^2$ profiles close to the edge of the IBL, plotted on the same abscissa. In both figures, shading of the symbols indicates the fetch with the colour scheme detailed in figure \ref{fig:x_d_color}. The four profiles in this figure are obtained at $\hat{x}/\delta_0 = 0.2,\,0.8,\,3.4\,$ and 15.7. Note that turbulence intensity profiles are shown only at every second streamwise location measured for clarity. The symbols with thick black outlines represent the location of the edge of the IBL. The solid horizontal line in (\textit{b}) is at $\Updelta(\overline{u^2}/U_{\infty}^2)/\Updelta\log_{10}(\hat{x}/\delta_0)  = 0$, and the dashed line shows the threshold $\Updelta(\overline{u^2}/U_{\infty}^2)/\Updelta\log_{10}(\hat{x}/\delta_0) = 1\times10^{-3}$.}
    \label{fig:uvar_IBL}
\end{figure}

The extent of flow recovery can also be quantified by the growth of the IBL. The IBL height $\delta_i$ at each streamwise location is calculated based on the difference between the $\overline{u^2}/U^2_{\infty}$ profile at the current location and the neighbouring upstream measurement location, i.e. $\delta_i$ is defined as the wall-normal location where $\partial (\overline{u^2}/U^2_{\infty})/\partial x \rightarrow 0$. It is well-known that in a canonical boundary layer with no surface heterogeneity, $\overline{u^2}$ exhibits outer-layer similarity only when normalised by $U_{\tau}^2$, and a dependence on Reynolds number presents if the velocity scale $U_{\infty}$ is used instead. However, the adequacy of this approach can be justified considering that the largest change of Reynolds number between the neighbouring profiles used to compute $\delta_i$ is usually within $10\%$, and the difference in the Reynolds number is negligible close to the roughness transition. The majority of the measurements are concentrated in this region owing to the logarithmic streamwise spacing employed for these measurements. It has been shown that $\delta_i$ determined from the turbulence intensity profile is comparable with the results from the more conventional methods based on the mean velocity profiles \citep{pendergrass1984dispersion,rouhi2018,MogengJFM2019}. Here we favour the turbulence intensity approach as the distinction associated with the roughness change is more pronounced in $\overline{u^2}$ compared to $U$ and less subject to small uncertainties in the measurement, resulting in a more robust estimation of $\delta_i$.  Figure \ref{fig:uvar_IBL} illustrates the process of extracting $\delta_i$ from the outer-scaled turbulence intensity profiles. Figure \ref{fig:uvar_IBL}(\textit{a}) shows good collapse in the outer layer with no appreciable Reynolds number trend, and the decrease in the turbulence intensity related to the internal layer growth is much more pronounced compared to the negligible Reynolds number trend. Figure \ref{fig:uvar_IBL}(\textit{b}) shows the difference between two consecutive outer-scaled turbulence intensity profiles $\Updelta (\overline{u^2}/U^2_{\infty})$ divided by $\Updelta \log_{10}(\hat{x}/\delta_0)$, the difference in the logarithmic fetch of these two streamwise locations. We use this difference to extract the IBL height $\delta_i$. In practice, a threshold of $\Updelta (\overline{u^2}/U^2_{\infty})/\Updelta\log_{10}(\hat{x}/\delta_0)  = 1\times 10^{-3}$ rather than 0 is selected to account for the noise in measurements. Note that there is also a weak Reynolds number trend in $\overline{u^2}/U^2_{\infty}$ profiles with increasing fetch: considering a rough-wall turbulent boundary layer satisfying the outer-layer similarity hypothesis, i.e. $\overline{u^2}/U^2_{\tau} = f(z/\delta_{99})$, the streamwise difference $\Updelta(\overline{u^2}/U^2_{\infty})/\Updelta\log_{10}(\hat{x}/\delta_0) =f(z/\delta_{99})\Updelta C_f/(2\Updelta\log_{10}(\hat{x}/\delta_0)) $, which is around $3\times10^{-4}$ or less. Therefore, with the current approach, the weak Reynolds number trend has insignificant effect on the $\delta_i$ results, especially in the near field. To examine the sensitivity of the resulting $\delta_i$ to the threshold, we reprocess the data of case \texttt{Re07ks16} with a threshold of $2\times10^{-3}$. Doubling the threshold leads to an underestimation up to 15\% in $\delta_i$ compared to that with a threshold of $1\times10^{-3}$. The exponent of a power-law fit through the data (to be discussed below) is changed from 0.77 to 0.75.  In summary, doubling the threshold leads to a smaller $\delta_i$, but the general trend of the data remains little changed. This method of extracting $\delta_i$ tends to pick up the upper edge of the modified region. In the vicinity of the rough-to-smooth change ($\hat{x}/k\lesssim10$), this method could presumably detect the upper limit of the roughness trailing wake (see figure \ref{fig:IBL_sketch}).

\begin{figure}
    \centering
	\setlength{\unitlength}{1cm}
	\begin{picture}(13,5.1)
      \put(0,0){\includegraphics[scale = 0.92]{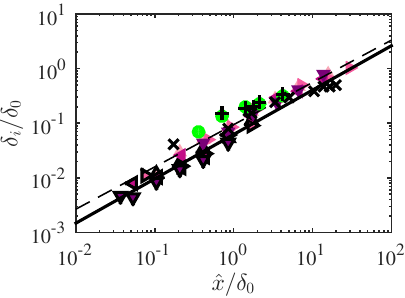}}
      \put(6.5,0){\includegraphics[scale = 0.92]{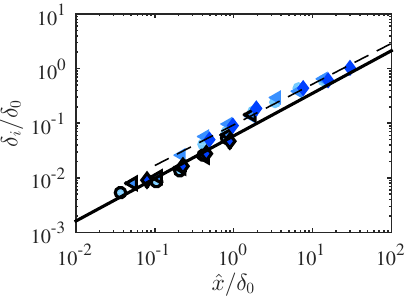}}
	\put(0.1,4.8){(\textit{a})}
	\put(6.6,4.8){(\textit{b})}
	\end{picture}

    \caption{IBL thickness $\delta_i$ versus the fetch $\hat{x}$ over the downstream smooth surface, both normalised by $\delta_0$, the boundary layer thickness at the roughness transition (coloured symbols without outlines). The symbols with a thick black outline are $\delta_i$ calculated using the same definition as in \citet{antonia1972response}. The colour of the symbols indicates (\textit{a}) $Re_{\tau 0}$ and (\textit{b}) $k_{s0}^+$, as defined in table \ref{tab:cases}. The thin dashed line is equation (\ref{Eq:IBL}), a power-law fit through $\delta_i$ calculated from the variance profile using the present method, and the thick solid line is a power-law fit through $\delta_i$ calculated using the definition of \citet{antonia1972response}. For comparison, $\delta_i$ reported in the original study of \citet{antonia1972response} is shown in (\textit{a}), where the black pluses and green dots represent $\delta_i$ determined as the inflection point in the $U$ versus $z^{1/2}$ profile and the `merging point' in the neighbouring mean velocity profiles where $\Updelta U=0$, respectively. The black crosses are $\delta_i$ determined from the $U$ versus $z^{1/2}$ profile of the grit case in \cite{Hanson2016}.}
    \label{fig:IBL}
\end{figure}

Using the method described above, $\delta_i$ at various streamwise locations is calculated for all cases and presented in figure \ref{fig:IBL}. Both $\delta_i$ and $\hat{x}$ are normalised by $\delta_0$, the boundary layer thickness at the rough-to-smooth transition. All data points collapse on to a straight line in logarithmic scale with no distinguishable trend with $Re_{\tau 0}$ or $k_{s0}^+$. A power-law fit
\begin{equation} 
\delta_i/\delta_{0} = A_0 (\hat{x}/\delta_0)^{b_0}
\label{Eq:IBL}
\end{equation}
results in coefficients of $A_0 = 0.094$ and $b_0 = 0.77$ for \emph{Group-Re}, and $A_0 = 0.095$ and $b_0 = 0.75$ for \emph{Group-ks}. This agrees closely with the observations of \citet{Bradley1968} and \citet{mulhearn1978wind}, where $\delta_i$ is defined as the `merging point' in the mean velocity profile. The growth appears to be more aggressive than $\delta_i\propto\hat{x}^{0.43}$ as reported by \citet{antonia1972response}. The major differences are: in their study, (1) $\delta_i$ is defined as the inflection point in the $U$ versus $z^{1/2}$ plot and (2) the upstream surface is roughened by 2D square ribs instead of sandpaper. In order to further clarify the underlying reason behind the disparity between these two results, we present $\delta_i$ estimated from the present dataset using the same definition as in \citet{antonia1972response}. The resulting $\delta_i$ is shown in figure \ref{fig:IBL} by symbols with the corresponding shape and colour but with a thick black outline. Note that we only show the data points where $\delta_i/\delta_{99}<0.15$ is satisfied to eliminate the contamination from the wake profile in this method. A power-law fit through the data points in each group is also shown in the figure by a thick solid line. The fitting coefficients are $A_{0} = 0.062$ and $b_{0}  = 0.81$ for \emph{Group-Re}, and $A_{0}  = 0.059$ and $b_{0}  = 0.78$ for \emph{Group-ks}. Thus the $\delta_i$ calculated using Antonia and Luxton's definition appears to be lower than the results obtained by thresholding the variance profile. This is expected as the former definition returns the `mid-point' of a dispersed or fluctuating IBL, while the current definition is more likely to pick up its upper limit. Regardless, exponents from the power-law fits of both \emph{Group-Re} and \emph{Group-ks} using Antonia \& Luxton's definition for $\delta_i$  are approximately 0.8, which is very close to the result using the current variance-based definition. In fact, the power-law exponent itself is a very sensitive quantity to assess. The streamwise extent where the fit is performed, the noise and scatter in the data points, and the step height effect at small fetches (see \citet{itt2018}, for example) can all affect the resulting power-law exponent. As shown by the green symbols in figure \ref{fig:IBL}(\textit{a}), the $\delta_i$ versus $\hat{x}$ trend in \cite{antonia1972response} is very similar to that in the present study. Note that in \cite{antonia1972response}, the data point at $\hat{x}/\delta_0 \approx 0.4$ ($\hat{x} = 1\mathrm{in}$) is omitted from the fit. A fit through all available data points will potentially bring the exponent closer to 0.8. In addition, the black crosses in figure \ref{fig:IBL}(\textit{a}) are $\delta_i$ of the grit case in \cite{Hanson2016}, which is also determined following Antonia \& Luxton's definition. They are observed to follow the trend in the current data except for the undershoot in the far field, where $\delta_i$ is presumably in the wake region. To conclude, the power-law exponents of $\delta_i$ determined from both variance and $U$ versus $z^{1/2}$ profiles are very similar, although the latter gives a slightly lower multiplicative coefficient $A_0$ as dictated by the nature of the method. The growth trends of $\delta_i$ in the current study are comparable with the past works, including those with a seemingly different power-law exponent. The underlying reason is that the power-law exponent is very sensitive to noise or uncertainty in $\delta_i$, especially when the data points are few, or the power law fit is conducted over a limited range of $\hat{x}/\delta_0$.

\begin{figure}
\centering
\setlength{\unitlength}{1cm}
	\begin{picture}(13,5.1)
\put(0,0){\includegraphics[scale = 0.95]{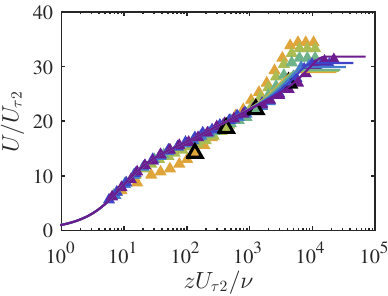}}
\put(6.5,0){\includegraphics[scale = 0.95]{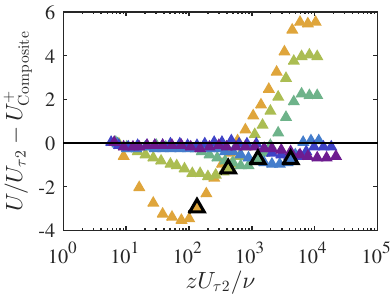}}
\put(0.1,4.8){(\textit{a})}
\put(6.6,4.8){(\textit{b})}
\end{picture}
\caption{(\textit{a}) Inner-normalised mean velocity profile for case \texttt{Re07ks16}. The solid coloured lines are composite profiles computed at matched $Re_{\tau}$ following \citet{Chauhan2009}. (\textit{b}) The difference between the measured mean streamwise velocity $U$ normalised by the local friction velocity $U_{\tau 2}$ and the composite profile $U^+_{\mathrm{Composite}}$ for the same case. In both figures, shading of the symbols changes from yellow to green and then to violet as the fetch increases. The colour scheme is summarised in detail in figure \ref{fig:x_d_color}. The profiles shown in the figure are obtained at $\hat{x}/\delta_0 = 0.2,\,0.8,\,3.4,\,15.7,\,39.4$ and 112. The symbols with thick black outlines show the location of the IBL (detailed in \S\ref{sec:big_IBL}) in each profile. Note that the mean velocity profiles are shown only at every second streamwise location measured for clarity.}
\label{fig:mean_U}
\end{figure}

Based on the fitted power law, the downstream fetch required for the IBL to reach the local $\delta_{99}$ is estimated to be $\hat{x}/\delta_0 = 26.5$, which is at approximately the same location as where the maximum of $C_f$ occurs (see figure \ref{fig:Cf}\textit{a},\textit{b}). However, we would like to re-emphasise that here we adopt the definition of the IBL as the region where the flow is modified by the new wall condition, and the flow within the IBL has been shown to be in a non-equilibrium state with the local wall conditions \citep[see][]{Antonia1971a, rouhi2018, MogengJFM2019}. This implies that even when $\delta_i \rightarrow \delta_{99}$ (when the internal layer has grown to the full layer height), the boundary layer may still not be in full quasi-equilibrium with the new wall condition. A complete recovery of all flow statistics to the quasi-equilibrium state is expected at a longer fetch, as will be evidenced in the following sections.

\section{Mean streamwise velocity}\label{sec:big_mean}

We utilise the hotwire data to further investigate the recovery of the mean flow statistics. We will primarily present the results of case \texttt{Re07ks16} as it permits the greatest streamwise development downstream of the roughness transition (hence demonstrating the asymptotic trends more clearly), however, very similar trends are observed in other cases and are not shown here for brevity. Mean streamwise velocity profiles normalised by the local $U_{\tau 2}$ over the smooth wall measured using OFI are shown in figure \ref{fig:mean_U}(\textit{a}). The coloured lines in the figure are composite velocity profiles at matched $Re_{\tau}$, following the expression of \citet{Chauhan2009}. They represent a reference at a quasi-equilibrium state with the smooth surface. The measured mean velocity profile overshoots the corresponding composite profile in the wake region, and this overshoot (the wake strength) diminishes with the downstream fetch as dictated by the recovering $C_f$ (see figure \ref{fig:Cf}), approaching the smooth-wall reference. The recovery of the flow at different wall-normal locations occurs at different rates: as shown in figure \ref{fig:mean_U}(\textit{a}), the mean velocity profile in the buffer region conforms to the smooth-wall reference after a short fetch of $\hat{x}/\delta_0=3.4$, while it takes $\hat{x}/\delta_0\gtrsim20$ for the wake region to recover.

The locations of the IBL in each profile as determined in \S\ref{sec:big_IBL} are shown by the symbols with thick black outlines. The measured mean velocity profiles exhibit a good collapse with the reference composite profile at $z^+<10$, however, there is a general lack of agreement in the buffer region and beyond, despite the fact that this region is well below the edge of the IBL. This confirms that the flow within the IBL is not in equilibrium with the local wall condition \citep{antonia1972response, ismail2018simulations, rouhi2018, MogengJFM2019}. The deviation from the smooth-wall composite profile is better demonstrated by viewing the velocity difference $U/U_{\tau 2}-U^+_{\mathrm{Composite}}$ shown in figure \ref{fig:mean_U}(\textit{b}). The difference between the measured profile and the smooth-wall reference persists even beyond $\hat{x}/\delta_0>10$ in the logarithmic region (see light blue symbols in figure \ref{fig:mean_U}\textit{b}). 

\begin{figure}
\centering
\setlength{\unitlength}{1cm}
\begin{picture}(13,5.1)
\put(0,0){\includegraphics[scale = 0.95]{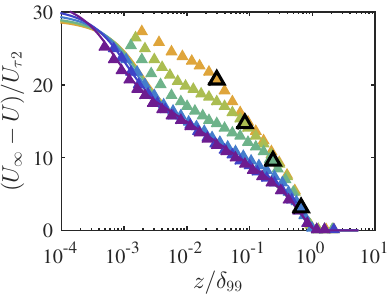}}
\put(6.5,0){\includegraphics[scale = 0.95]{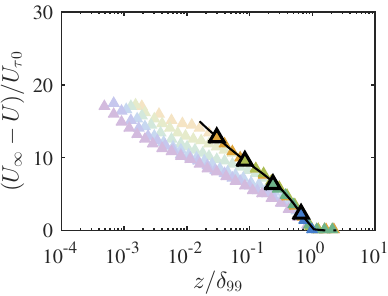}}
\put(0.1,4.8){(\textit{a})}
\put(6.6,4.8){(\textit{b})}
\end{picture}
\caption{Mean velocity deficit $U_{\infty}-U$ for case \texttt{Re07ks16} normalised by (\textit{a}) $U_{\tau 2}$, the local friction velocity obtained on the smooth wall and (\textit{b}) $U_{\tau 0}$, the friction velocity just upstream on the roughness transition over the rough surface. The abscissa is the outer-scaled wall-normal location $z/\delta_{99}$. The coloured lines in (\textit{a}) are reference smooth-wall composite profiles with $Re_{\tau}$ matched at each streamwise measuring locations, and the solid black line in (\textit{b}) is the reference rough-wall hotwire profile obtained in the present study, with the datum adjusted to the smooth surface. The shading of the symbols indicates the downstream fetch, with the specific streamwise locations detailed in the caption of figure \ref{fig:mean_U}. Data points within the IBL are shown with a lighter tint in (\textit{b}) to highlight the outer-layer behaviour. Note that in the two most downstream locations (dark blue and violet symbols), $\delta_i$ has already reached $\delta_{99}$, therefore the entire profile is shown in the lighter tint and no IBL location is marked.}
\label{fig:U_defect}
\end{figure}

The mean velocity deficit profiles normalised by the friction velocity $U_{\tau 2}$ obtained locally at the wall are shown in figure \ref{fig:U_defect}(\textit{a}). The corresponding smooth-wall composite profiles at matched $Re_{\tau}$ are again shown by coloured lines. A collapse on the smooth-wall reference is observed in the outer layer only at the most downstream locations (at $\hat{x}/\delta_0 = 39.4$ and 112, as shown by the dark blue and violet symbols) where the flow has completely recovered to an equilibrium state with the smooth-wall condition. The deficit profiles closer to the step change, however, are above the reference profile, and they gradually relax back to the reference as the fetch increases. The lack of collapse in the inner-normalised velocity deficit suggests an alternative velocity scale in the outer layer above the IBL. We assume the outer-layer velocity scale $U_{\tau,\mathrm{out}} = U_{\tau 0}$ as a constant, and present the velocity deficit profiles normalised by this velocity scale in figure \ref{fig:U_defect}(\textit{b}). Under this scaling, the velocity deficit profiles above the IBL collapse reasonably well on the rough-wall reference, confirming that the flow beyond the IBL scales on the upstream characteristic scales. More details on the evolution of the outer layer can be found in the appendix \S\ref{sec:big_outer}. In essence, for the first few boundary layer thicknesses downstream of the roughness transition, the velocity and length scale in the outer layer evolve with increasing $\hat{x}/\delta_0$ in a similar fashion as if the roughness transition did not exist and the upstream roughness extended beyond $\hat{x}>0$. Note that the rough-wall reference deficit profile is essentially the same as the smooth-wall deficit profile under outer scaling, as demonstrated in figure \ref{fig:Rough_OutSim}(\textit{a}). For the two most downstream locations (dark blue and violet symbols), $\delta_i$ has reached $\delta_{99}$, therefore no collapse with the rough-wall reference is found under outer scaling. 

For figure \ref{fig:U_defect}(\textit{b}), the IBL thicknesses (shown by the symbols with thick black outlines), although determined from the streamwise turbulence intensity profiles, approximately coincide with the location where the mean velocity deficit profile deviates from the rough-wall reference. This is also true for other cases in the current study and the figures are not shown here for brevity. This implies that for the range of $Re_{\tau 0}$ and $k_{s0}^+$ investigated, the IBL thickness determined from the mean streamwise velocity and the streamwise turbulence intensity profiles are similar. A similar observation has also been reported by \cite{pendergrass1984dispersion} in an experimental study of an internal layer formed over a rough-to-rougher transition, as well as in an open-channel DNS study with a rough-to-smooth transition \citep{rouhi2018} where the intersection of two logarithmic laws in the mean velocity profile coincides with the wall-normal location where the turbulence intensity profile deviates from its upstream counterpart. 

\section{Blending model of the mean velocity profile} \label{sec:blending}

\begin{figure}
    \centering
    \includegraphics[scale = 0.65]{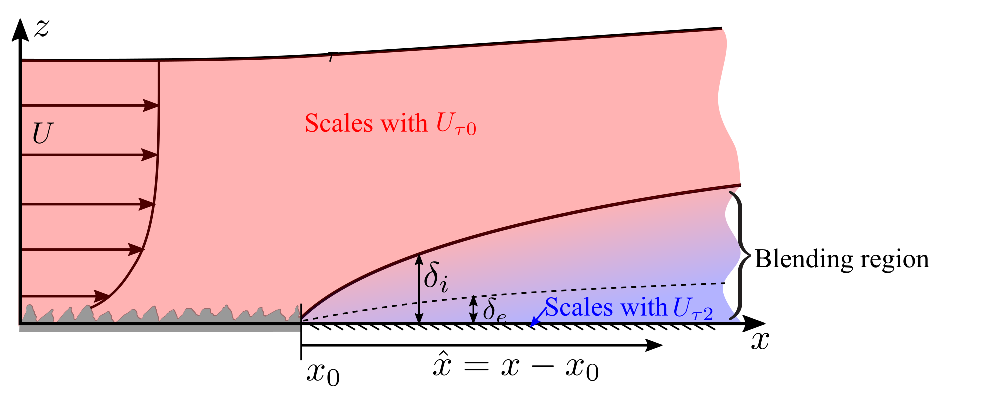} 
    \caption{Sketch of a turbulent boundary layer over a rough-to-smooth change, featuring a gradual adjustment of the flow from smooth-wall behaviour (blue) close to the wall to rough-wall behaviour (red) deeper into the flow. The flow is from left to right.}
    \label{fig:Blend_sketch}
\end{figure}

In the seminal work of \citet{Elliott1958}, a theoretical model of a turbulent boundary layer adapting to a step change in the surface condition was developed by evolving an assumed mean velocity profile in the streamwise direction using the momentum equation. The flow both within and above the IBL is modelled by a logarithmic law with the friction velocity chosen as $U_{\tau 2}$ and $U_{\tau 0}$, respectively, representing the stress jump that is assumed to occur at the edge of the IBL. As shown by the experimental data in the previous section, the mean velocity profile above the IBL can be approximated by the upstream rough-wall profile with $U_{\tau 0}$ as the velocity scale, while it deviates from the corresponding smooth-wall reference above the equilibrium layer. The deviation of the mean velocity profile from a smooth-wall profile in the IBL has also been previously reported \citep[][for example]{Antonia1971a, Garratt1990, ismail2018simulations, rouhi2018, MogengJFM2019}, demonstrating the necessity of improving the classical mean velocity profile model. In this section, we develop a semi-empirical model of the mean velocity profile based on the understanding gained using the present experimental data. 

\subsection{Description of the model}
\label{sec:descrip_blending}
A schematic of the proposed model is illustrated in figure \ref{fig:Blend_sketch}. Downstream of a rough-to-smooth change, the flow very close to the wall (within the EL) is expected to have fully adapted to the new wall condition, so it can be approximated by a canonical smooth-wall profile. Above the IBL, the flow is expected to follow the upstream rough-wall scaling and is approximated by a rough-wall profile. We use blue and red colours, respectively, to represent the completely smooth-wall and rough-wall states of the flow. With an increasing $z$, we expect the mean flow to gradually transition from the smooth-wall asymptote and be increasingly similar to the rough-wall one, as illustrated by the transition from blue to red in figure \ref{fig:Blend_sketch}.

A mathematical description of the qualitative model outlined above is detailed as follows. In the near-wall region where the flow is in full equilibrium with the local smooth wall condition (shown by the blue coloured region in figure \ref{fig:Blend_sketch}), the mean velocity distribution satisfies
\begin{equation}
\frac{U(z)}{U_{\tau 2}} = f(z^+) \equiv \frac{U_S(z_S)}{U_{\tau S}}, \quad (z\rightarrow 0)
\label{eq:blend_S}
\end{equation}
where $U_{\tau 2}$ is the local friction velocity measured by OFI on the smooth wall, $z^+ = zU_{\tau 2}/\nu$ for the rough-to-smooth profile and $z^+ = z_{S}U_{\tau S}/\nu$ for the smooth-wall reference. The subscript $(\cdot)_S$ denotes quantities of the smooth-wall reference. Similarly, above the IBL, the outer-layer similarity is satisfied when normalised by the correct velocity scale, of which $U_{\tau 0}$ has been found to be a good approximation when the fetch is not too large (see \S\ref{sec:big_mean}). The region where the upstream rough-wall scaling is supposed to hold is shown by the red patch in figure \ref{fig:Blend_sketch}. The velocity deficit in the outer layer can be expressed as
\begin{equation}
\frac{U_{\infty} - U(z)}{U_{\tau 0}} = g(\eta) \equiv \frac{U_{\infty} - U_{R}(z_R)}{U_{\tau 0}}, \quad (z>\delta_i)
\label{eq:blend_R}
\end{equation}
where $\eta$ is the wall-normal distance normalised by the boundary layer thickness of the corresponding velocity profile, i.e. $\eta = z/\delta_{99}$ for the rough-to-smooth profile and $\eta = z_{R}/\delta_{0}$ for the rough-wall reference. The subscript $(\cdot)_R$ denotes quantities of the rough-wall reference. Note that here the velocity scale is selected as $U_{\tau 0}$, the rough-wall friction velocity, for both upstream and downstream mean velocity profiles, as there is no apparent change in the outer-layer velocity scale after the rough-to-smooth change (see \S\ref{sec:big_outer}). (\ref{eq:blend_R}) essentially reduces to $U(\eta) = U_R(\eta)$, therefore a rough-wall velocity profile can be recovered above the IBL (which will include a log and a wake region). As required by (7.2), the applicability of the model relies on an outer-layer similarity above the IBL. For extreme scenarios (such as a very small $\delta_0/k$ ratio) where the outer-layer similarity no longer holds, the model would not be expect to work.

\begin{figure}
    \centering
\setlength{\unitlength}{1cm}
\begin{picture}(13,8.6)
\put(0,0){\includegraphics[scale = 0.905]{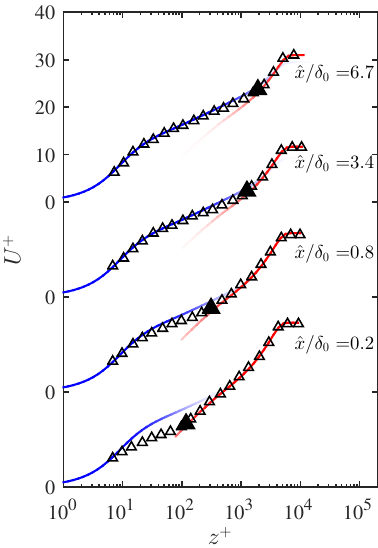}}
\put(6.5,0){\includegraphics[scale = 0.905]{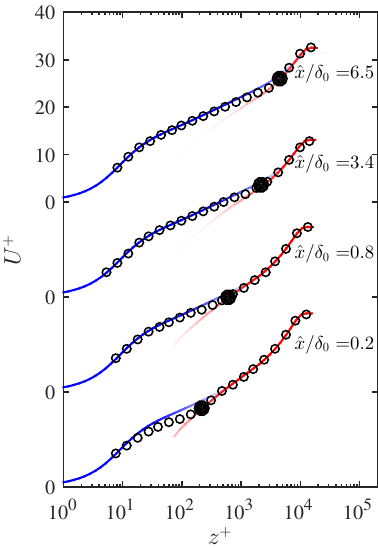}}
\put(0.1,8.3){(\textit{a})}
\put(6.6,8.3){(\textit{b})}
\end{picture}
    \caption{Mean velocity profiles downstream of a rough-to-smooth change normalised by $U_{\tau 2}$ (empty symbols) compared to the rough-wall and smooth-wall profiles. The blue line is $U_S^+$ and the red line is $U_R^{*+}$. The solid black symbol represents the location of $\delta_i$ extracted from the turbulence intensity profile as detailed in \S\ref{sec:big_IBL}. (\textit{a}) is from case \texttt{Re07ks16} and (\textit{b}) is from case \texttt{Re14ks11}. }
    \label{fig:Blend_U}
\end{figure}

In order to combine the two limiting cases described in (\ref{eq:blend_S}) and (\ref{eq:blend_R}) and demonstrate how the recovering profile transitions from the smooth-wall to the rough-wall asymptote as the wall-normal distance increases, we define
\begin{equation}
U^{*+}_{R} = \frac{U_{R}(z_R)}{U_{\tau 2}},
\label{eq:blend_UR}
\end{equation}
and plot it against
\begin{equation}
z^{*+}_{R} = \eta\frac{\delta_{99} U_{\tau 2}}{\nu}= \frac{\delta_{99}}{\delta_{0}}\frac{z_R U_{\tau2}}{\nu}
\label{eq:blend_zR}
\end{equation}
on the inner-normalised mean velocity profiles as shown in figure \ref{fig:Blend_U}. The $U^{*+}_{R}$ versus $z^{*+}_R$ profile is designed to match the viscous-scaled $U^+$ versus $z^+$ profile above the IBL through an assumed outer-layer similarity (\ref{eq:blend_R}), representing the outer layer which is scaled on with the upstream rough-wall velocity scale. The asterisk in the notation is to distinguish it from the conventional inner normalisation where the corresponding friction velocity $U_{\tau {R}}$ is used. The smooth-wall reference is chosen as a composite velocity profile \citep{Chauhan2009} with the same $Re_{\tau}$ as the rough-to-smooth profile, and the rough-wall reference is obtained from the hotwire survey just upstream of the roughness transition at matched flow conditions. 

Without losing generality, we express the recovering mean velocity profile as 
\begin{equation}
U^+ = U_S^+E+U_R^{*+}(1-E),
\label{eq:blend}
\end{equation}
following \cite{Krug2017}, who approximated statistics in the wake region of a boundary layer by blending the turbulent and non-turbulent components through a Gaussian distribution of the interface. Here, instead of prescribing a functional form of the blending function $E(z^+)$, we aim at obtaining that from the measured velocity profiles. At this stage, the only requirement of $E(z^+)$ is that it decreases monotonically from 1 to 0 when $z^+$ increases from 0 to $\delta_{i}^+$, corresponding to the limiting cases as the flow fully follows the smooth-wall profile in the vicinity of the wall, and it entirely preserves the upstream rough-wall scaling above the IBL. The EL is implicitly modelled in $E(z^+)$, which will be discussed at the end of this section.

\begin{figure}
    \centering
    \includegraphics[scale = 0.95]{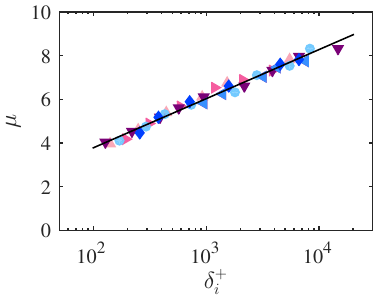} 
    \hspace{3mm}
    \includegraphics[scale = 0.95]{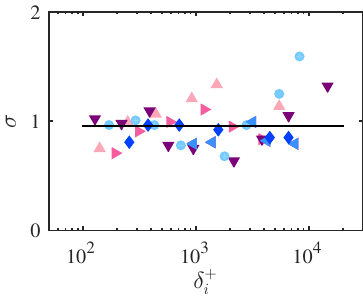} 
	\put(-372,140){(\textit{a})}
	\put(-175,140){(\textit{b})}
    \caption{(\textit{a}) The mean $\mu$ and (\textit{b}) standard deviation $\sigma$ of all cases obtained by fitting (\ref{eq:blend_log}) to $E(z^+)$ determined experimentally from (\ref{eq:blend_frac}). The colour of the symbols indicates $Re_{\tau 0}$ or $k_{s0}^+$ as defined in table \ref{tab:cases}. Expressions for the solid lines in the figures are given by (\ref{eq:blend_mu_sigma}).}
    \label{fig:Blend_mu_sigma}
\end{figure}

\begin{figure}
    \centering
    \includegraphics[scale = 0.95]{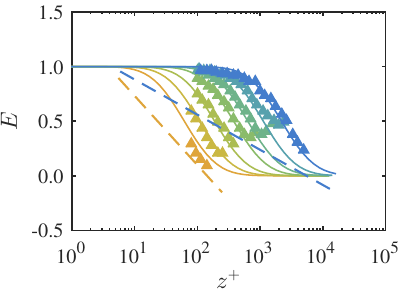} 
    \hspace{3mm}
    \includegraphics[scale = 0.95]{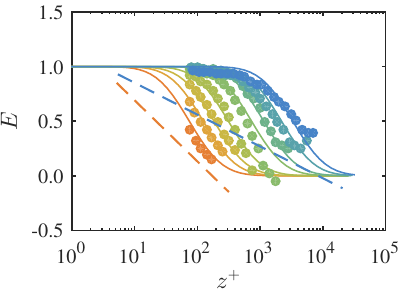} 
\put(-395,140){(\textit{a})}
\put(-190,140){(\textit{b})}\\
	\includegraphics[scale = 0.95]{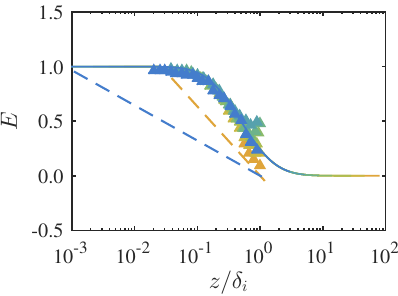} 
    \hspace{3mm}
    \includegraphics[scale = 0.95]{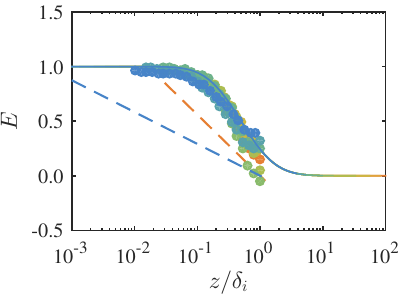} 
	\put(-395,140){(\textit{c})}
	\put(-190,140){(\textit{d})}

    \caption{The blending function $E$ for (\textit{a}, \textit{c}) \texttt{Re07ks16} and (\textit{b}, \textit{d}) \texttt{Re14ks11}. Shading indicates the downstream fetch: $\hat{x}$ increases as the colour changes from yellow to blue in each case, as defined in figure \ref{fig:x_d_color}. The symbols are computed from each hotwire profile from the wall up to the edge of the IBL following (\ref{eq:blend_frac}), and the solid lines with corresponding colours represent $E$ reconstructed using (\ref{eq:blend_log}) with the parameters given by (\ref{eq:blend_mu_sigma}). The dashed lines are (\ref{eq:blend_Chamorro}) at corresponding streamwise locations, as proposed by \cite{Chamorro2009}. (\textit{c}) and (\textit{d}) are the original and reconstructed $E$ plotted against $z/\delta_i$.}
    \label{fig:Blend_Gamma}
\end{figure}

The blending function can be derived by restating equation (\ref{eq:blend}) as
\begin{equation}
E = \frac{U^+-U_R^{*+}}{U_S^+-U_R^{*+}}.
\label{eq:blend_frac}
\end{equation}
Here $U_{S}^+$ is the equilibrium smooth-wall composite profile, $U_R^+$ comes from the measured rough-wall profile upstream of the transition, and we have measured $U^+$ at multiple locations downstream of the transition. The experimentally obtained blending function $E(z^+)$ is shown by the symbols in figure \ref{fig:Blend_Gamma} for cases \texttt{Re07ks16} (\textit{a}) and \texttt{Re14ks11} (\textit{b}) as an example.

\begin{figure}
    \centering
\setlength{\unitlength}{1cm}
\begin{picture}(13,17.2)
\put(0,0){\includegraphics[scale = 0.905]{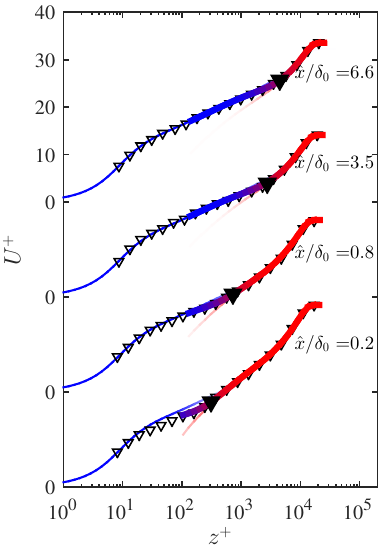}}
\put(6.5,0){\includegraphics[scale = 0.905]{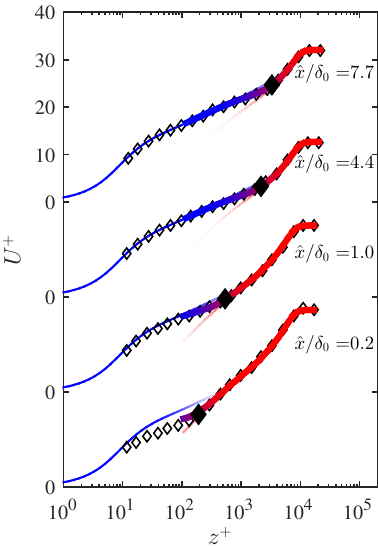}}
\put(0,8.6){\includegraphics[scale = 0.905]{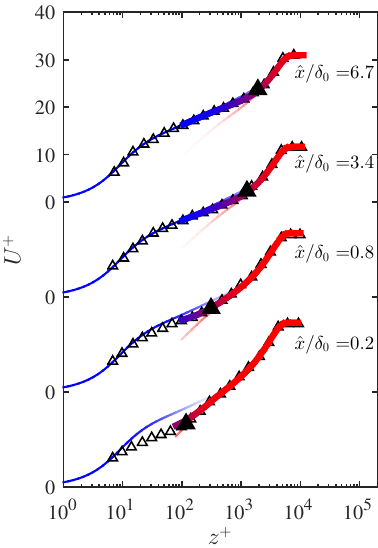}}
\put(6.5,8.6){\includegraphics[scale = 0.905]{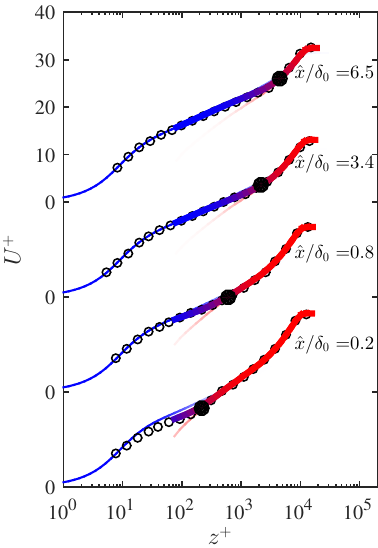}}
\put(0.1,16.9){(\textit{a})}
\put(6.6,16.9){(\textit{b})}
\put(0.1,8.3){(\textit{c})}
\put(6.6,8,3){(\textit{d})}
\end{picture}
    \caption{Comparison of the measured (empty symbols) and reconstructed (thick lines with colour transitioning from blue to red) mean velocity profiles downstream of a rough-to-smooth change.  The reconstructed velocity profile is obtained using (\ref{eq:blend_frac}) and (\ref{eq:blend_log}) with parameters given by (\ref{eq:blend_mu_sigma}). The blue line is the equilibrium smooth-wall profile $U_S^+$ and the red line is $U_R^{*+}$. The solid black symbol represents the location of $\delta_i$ extracted from the turbulence intensity profile as detailed in \S\ref{sec:big_IBL}. (\textit{a}) \texttt{Re07ks16}; (\textit{b}) \texttt{Re14ks11}; (\textit{c}) \texttt{Re21ks16}; (\textit{d}) \texttt{Re14ks22}. }
    \label{fig:Blend_U_recon}
\end{figure}

The experimentally measured behaviour of $E$ here appears to be well captured by the cumulative distribution function of a log-normal distribution:
\begin{equation}
E(z^+;\mu,\sigma) = \frac{1}{2}\left[1-\mathrm{erf}\left(\frac{\ln(z^+)-\mu}{\sqrt{2\sigma^2}}\right)\right],
\label{eq:blend_log}
\end{equation}
as shown by the solid lines in figure \ref{fig:Blend_Gamma}. The mean $\mu$ and standard deviation $\sigma$ obtained by fitting (\ref{eq:blend_log}) for all cases are summarised in figure \ref{fig:Blend_mu_sigma}. As expected, the averaged location of the flapping interface (captured by $\mu$) increases as $\delta_i^+$ increases with an increasing downstream fetch (figure \ref{fig:Blend_mu_sigma}\textit{a}), while the standard deviation of the flapping interface ($\sigma$) appears to remain approximately constant for the entire range. Little $Re_{\tau0}$ or $k_{s0}^+$ dependence is observed from the data, and a general trend for all cases is found to be
\begin{equation}
\mu = \ln(\delta_i^+)-0.7, \quad \sigma = 1.0.
\label{eq:blend_mu_sigma}
\end{equation}
The solid coloured lines in figure \ref{fig:Blend_Gamma} show the profiles of $E$ obtained from equations (\ref{eq:blend_log}) and (\ref{eq:blend_mu_sigma}), which compare very well with the experimentally determined $E$ (shown by the symbols). Subsequently, the recovering mean velocity profile can be obtained by substituting the reconstructed $E$ into (\ref{eq:blend}). The results are shown by the thick lines with a colour transitioning from blue to red in figure \ref{fig:Blend_U_recon}, where they are compared to the experimental data (symbols) and the assumed equilibrium profiles (in red and blue). The reconstructed mean velocity profiles are in a better agreement with the experimental data compared to the equilibrium $U_S^+$ profiles. The improvement is especially notable close to the roughness transition. The modelled velocity profiles of cases \texttt{Re21ks16} and \texttt{Re14ks22}, which have the largest $Re_{\tau0}$ and $k_{s0}^+$ respectively from the current dataset, are also shown in figure \ref{fig:Blend_U_recon}(\textit{c}) and (\textit{d}). The model performance remains satisfactory in these extreme cases. The reconstructed mean velocity profile is relatively insensitive to the choice of $\sigma$. Varying $\sigma$ from 0.7 to 1.5 (which is the range of scatter in figure \ref{fig:Blend_mu_sigma}\textit{b}) leads to less than 3\% change in the reconstructed mean velocity. The uncertainty in $U_{\tau 2}$ affects the blending velocity mainly in the near-wall region. In fact, very close to the wall where $E\rightarrow1$, the blending velocity profile is identical to the local smooth-wall reference, therefore the error in the blending velocity profile would simply manifest as scaling the mean velocity with a biased $U_{\tau 2}$.

After substituting the fitted expressions (\ref{eq:blend_mu_sigma}) into (\ref{eq:blend_log}), it is found that the blending function reduces to a self-similar form:
\begin{equation}
E(z/\delta_i) = \frac{1}{2}\left[1-\mathrm{erf}\left(\frac{\ln(z/\delta_i)+0.7}{\sqrt{2}}\right)\right].
\label{eq:blend_ss}
\end{equation}
Here, $\delta_i$ is the characteristic length scale and $E$ is a function of one single variable, $z/\delta_i$. The existence of the self-similar behaviour of $E$ is further confirmed by plotting the measured $E$ against $z/\delta_i$. As shown in figure \ref{fig:Blend_Gamma}(\textit{c,d}), data from all streamwise locations in both cases collapse on to a single trend that follows (\ref{eq:blend_ss}) closely (shown by the solid line). Note that as $z/\delta_i$ approaches 1, the denominator $U_S^+-U_R^{*+}$ in (\ref{eq:blend_frac}) becomes increasingly small, which can lead to a higher error and scatter in $E$. Nevertheless, the agreement between the measured and reconstructed mean velocity profile is barely affected due to the small difference between $U_S^+$ and $U_R^{*+}$ in this region.

In the existing models in the literature where a wall-normal blending profile is considered, the equilibrium layer thickness $\delta_e$ is usually prescribed as a fraction of $\delta_i$. \cite{abkar2012new} used $\delta_e/\delta_i = 0.027$ and \cite{ghaisas2020predictive} found $\delta_e/\delta_i = 0.004$ and 0.055 optimum for a wind tunnel dataset \citep{Chamorro2009} and a field dataset \citep{Bradley1968}, respectively. These are in good agreement with previous $\delta_e$ definitions based on the mean velocity or shear-stress profiles in numerical studies \citep{Shir1972, Rao1974, rouhi2018}. In the context of the blending model, $\delta_e$ can also be derived from $E$ without any assumption about its dependence on $\delta_i$. For instance, by setting a threshold at $E=0.99$ (i.e. roughly $1\%$ difference between $U^+$ and $U_S^+$), an equilibrium layer thickness of $\delta_e = 0.05\delta_i$ results, and a threshold at $E=0.95$ leads to $\delta_e = 0.1\delta_i$.  Note that the EL thickness determined from either $E$ or some other flow statistic profiles is sensitive to the choice of the threshold, especially at a large $\hat{x}/\delta_0$ where the recovery is near complete. Therefore, the agreement between these EL thicknesses should only be treated qualitatively.

\cite{Chamorro2009} proposed a similar method to model the non-equilibrium velocity profile with a blending function rearranged to be consistent with notations in this study as
\begin{equation}
E_c(z^+) = 1-\ln\left(\cfrac{z^+}{k_{s0}^+\exp(-\kappa A'_{fr})}\cfrac{U_{\tau R}}{U_{\tau 2}}\right)\left/\ln\left(\cfrac{\delta_i^+}{k_{s0}^+\exp(-\kappa A'_{fr})}\cfrac{U_{\tau R}}{U_{\tau 2}}\right).\right.
\label{eq:blend_Chamorro}
\end{equation}
This blending function $E_c$ at the first and the last streamwise locations is shown in figure \ref{fig:Blend_Gamma} by the dashed lines with corresponding colours. Note that the equilibrium smooth-wall and rough-wall profiles were originally modelled by the logarithmic law only, therefore a fair comparison can only be drawn in the expected logarithmic region. A better agreement with the symbols is observed closer to the step change, while the predicted $E_c$ falls below the measured data points downstream, resulting in an over-estimation of the contribution from the rough-wall profile during blending. 

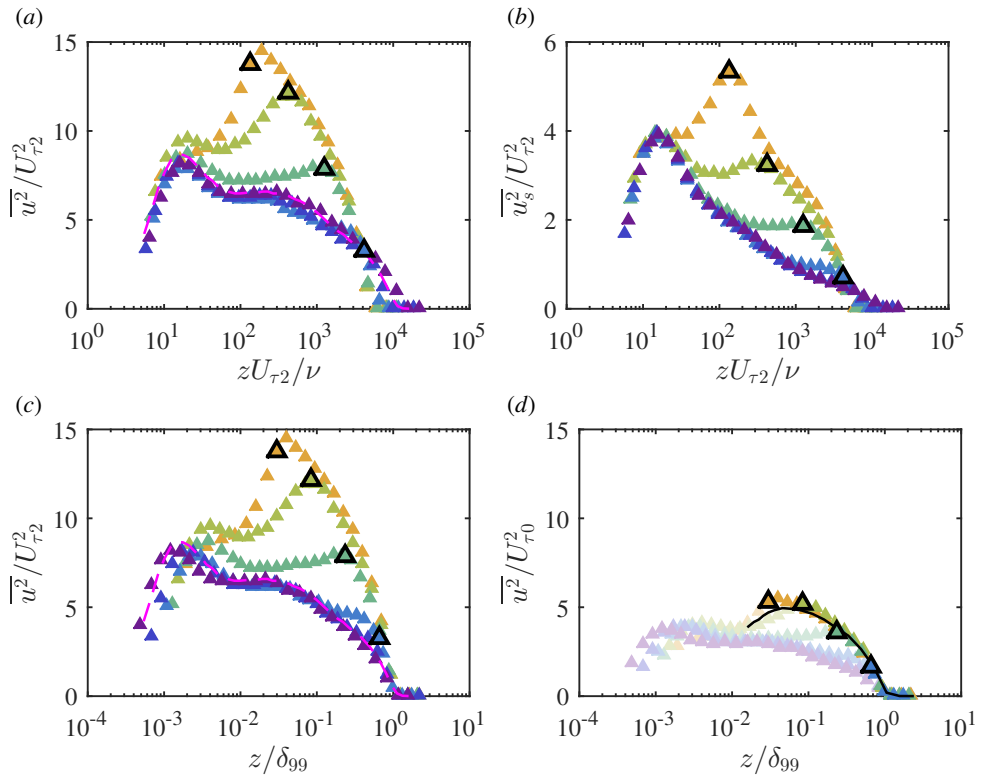
\begin{figure}
\centering
\setlength{\unitlength}{1cm}
\begin{picture}(13,10.2)
\put(0,5.1){\includegraphics[scale = 0.95]{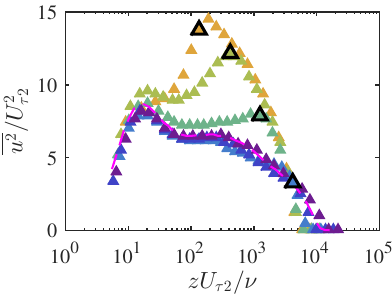}}
\put(6.5,5.1){\includegraphics[scale = 0.95]{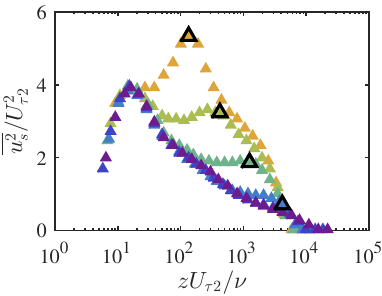}}
\put(0,0){\includegraphics[scale = 0.95]{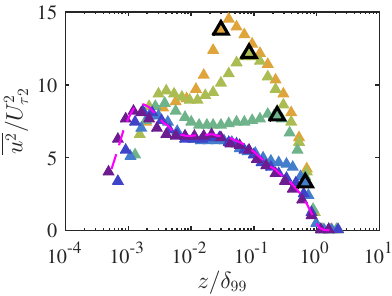}}
\put(6.5,0){\includegraphics[scale = 0.95]{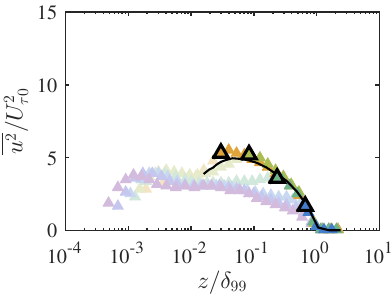}}
\put(0.1,9.9){(\textit{a})}
\put(6.6,9.9){(\textit{b})}
\put(0.1,4.8){(\textit{c})}
\put(6.6,4.8){(\textit{d})}
\end{picture}
\caption{(\textit{a}) Streamwise turbulence intensity $\overline{u^2}$ profiles and (\textit{b}) small-scale contribution to the broadband $\overline{u^2}$ normalised by the local $U_{\tau 2}$ and plotted against viscous scaled wall-normal distance. (\textit{c}) and (\textit{d}) are $\overline{u^2}$ normalised by the local $U_{\tau 2}$ and $U_{\tau0}$, respectively, and plotted against outer scaled wall-normal distance. All plots are for case \texttt{Re07ks16}. The symbols with thick black outlines show the location of the IBL (detailed in \S\ref{sec:big_IBL}) in each profile. The shading of the symbols indicates the downstream fetch, with the specific streamwise locations detailed in the caption of figure \ref{fig:mean_U}. The dashed magenta line in (\textit{a}) is a reference smooth-wall hotwire profile with $l^+=22.7$ measured at $Re_{\tau} = 1.0\times10^4$, a Reynolds number matched with the most downstream profile \citep{marusic2015evolution}. The solid black line in (\textit{d}) is the reference rough-wall hotwire profile, with $z = 0$ shifted downward by $|\Updelta H|$ (the distance between the roughness crest and the smooth wall) to the smooth surface. }
\label{fig:uvar}
\end{figure}

\section{Streamwise turbulence intensity}\label{sec:big_turb}

Viscous-scaled streamwise turbulence intensity profiles are shown in figure \ref{fig:uvar}(\textit{a}) for case \texttt{Re07ks16}. Close to the step change, the peak magnitude of $\overline{u^2}/U_{\tau 2}^2$ in the outer layer exceeds the inner peak at $z^+ = 15$ as a result of the highly turbulent rough-wall flow above the IBL, which imposes a footprint onto the near-wall region. The location of the `outer peak' coincides with the edge of the IBL (shown by the bold symbols), and the magnitude of this outer peak decreases with $\hat{x}$. A more gradual decreasing trend with $\hat{x}$ can also be observed at the inner peak, which eventually reaches the same value as a smooth-wall profile in equilibrium (dashed magenta line). In \cite{MogengJFM2019} over a limited range of $Re_{\tau 0}$ and $k_{s0}^+$, we suggested that the collapse in the small-scale energy could be used as a surrogate method of extracting the smooth-wall velocity scale $U_{\tau2}$ in situations where direct measurements in the viscous sublayer are not possible.  Figure \ref{fig:uvar}(\textit{b}) shows the small-scale contribution $\overline{u^2_s}/U_{\tau 2}^2$ to the broadband streamwise turbulence intensity. A threshold is selected as $1/f^+ = 200$ (to be detailed in \S\ref{sec:big_spectrum}), roughly equivalent to a streamwise wavelength of $\lambda_x^+ = 2000$ at the inner-peak location. Profiles at all streamwise locations collapse well at $z^+=15$, while the excess energy still presents close to and above $\delta_i$ (the IBL location is shown by the symbols with the bold outline). As noted previously, the small-scale contribution recovers almost immediately downstream of a roughness step change \citep{MogengJFM2019}, and the decreasing trend of the broadband turbulence intensity observed in figure \ref{fig:uvar}(\textit{a}) at the inner peak is due to the reduction in the footprint of the turbulent motions at larger scales caused presumably by the growth in $\delta_i$ with $\hat{x}$. This will be discussed more in depth in \S\ref{sec:big_spectrum}.

The outer ($z/\delta_{99}$) scaling of streamwise turbulence intensity profiles with both the local velocity scale ($U_{\tau 2}$) and the upstream rough-wall velocity scale ($U_{\tau 0}$) is examined in figures \ref{fig:uvar}(\textit{c}) and (\textit{d}), respectively. These plots can give some indication of the degree of outer layer similarity observed, but a lack of collapse in the near-wall region is enforced by the $z/\delta_{99}$ scaling as the abscissa. When normalised by $U_{\tau 2}$, the friction velocity obtained locally over the smooth wall, $\overline{u^2}/U_{\tau 2}^2$ downstream of the surface transition is much higher than the smooth-wall reference (shown by the dashed magenta line, which has a viscous-scaled hotwire length of $l^+=22.7$, comparable with the present dataset), as shown in \ref{fig:uvar}(\textit{c}). Only for the two most downstream locations ($\hat{x}/\delta_0 = 39.4$ and 112, shown by the dark blue and violet symbols) does the outer part of the variance profile scaled in this manner show signs of collapsing to the equilibrium smooth-wall reference. This is presumably because at $\hat{x}/\delta_0\geq39.4$, $\delta_i$ has grown close to the edge of the boundary layer, and the entire boundary layer has now almost fully recovered to the smooth wall condition, in consistency with the recovery of mean velocity profiles in figure \ref{fig:U_defect}. In figure \ref{fig:uvar}(\textit{d}), outer-layer similarity between the measurements downstream of the step change and the rough-wall reference (shown by the solid black curve) is observed beyond the IBL, when the upstream friction velocity is chosen as the velocity scale. Note that the dashed magenta line in figure \ref{fig:uvar}(\textit{c}) and the solid black line in figure \ref{fig:uvar}(\textit{d}) are expected to collapse under Townsend's assumption of outer-layer similarity, as previously demonstrated in figure \ref{fig:Rough_OutSim}(\textit{b}). This is similar to observations made in figure \ref{fig:U_defect} for the mean velocity deficit.

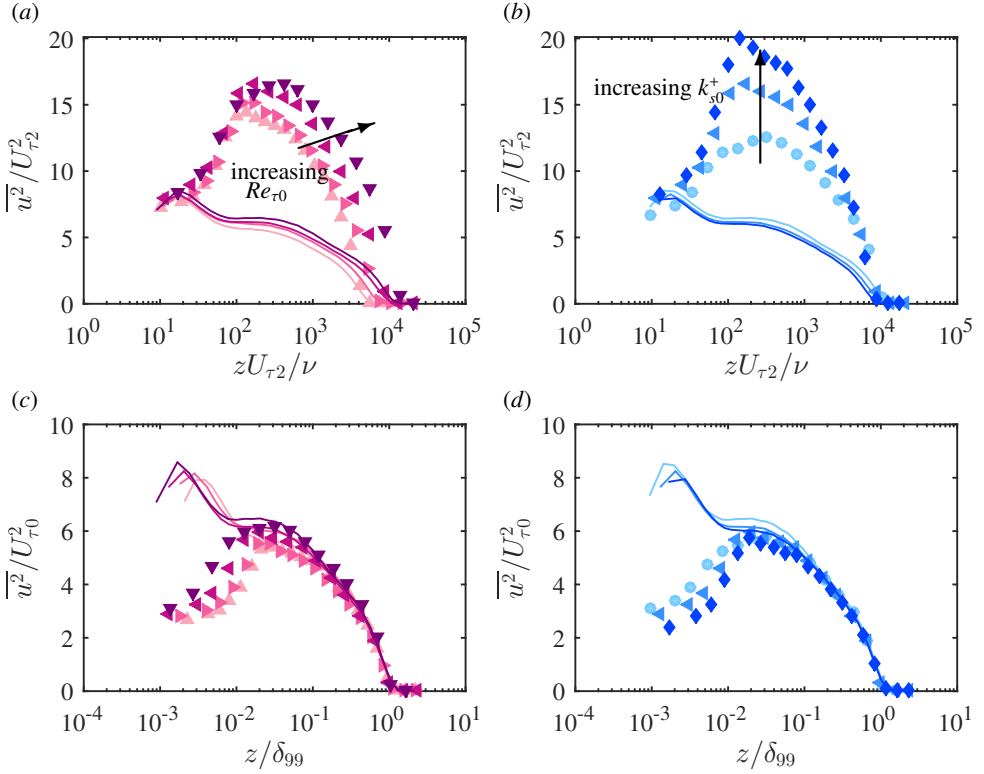
\begin{figure}
\centering
\setlength{\unitlength}{1cm}
\begin{picture}(13,10.2)
\put(0,5.1){\includegraphics[scale = 0.95]{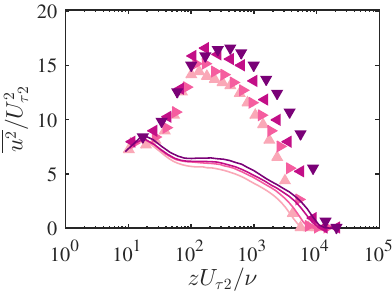}}
\put(6.5,5.1){\includegraphics[scale = 0.95]{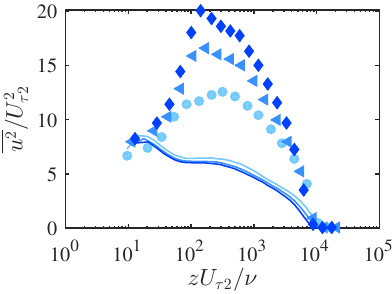}}
\put(0,0){\includegraphics[scale = 0.95]{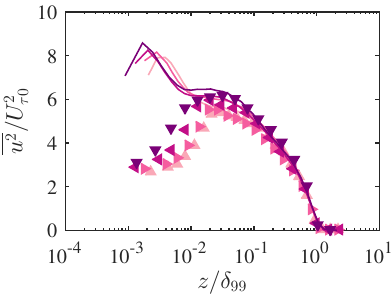}}
\put(6.5,0){\includegraphics[scale = 0.95]{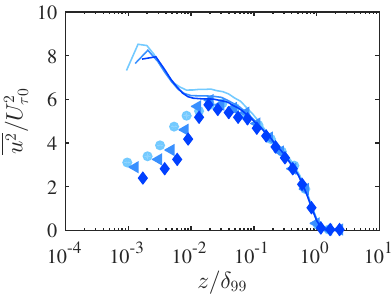}}
\put(0.1,9.9){(\textit{a})}
\put(6.6,9.9){(\textit{b})}
\put(0.1,4.8){(\textit{c})}
\put(6.6,4.8){(\textit{d})}
\thicklines
\put(3.9,8.2){\vector(3,1){1}}
\put(3,7.8){increasing}
\put(3.2,7.55){$Re_{\tau 0}$}
\put(10,8){\vector(0,1){1.5}}
\put(7.8,8.9){increasing $k_{s 0}^+$}
\end{picture}
\caption{Streamwise turbulence intensity $\overline{u^2}$ at $\hat{x}/\delta_0\approx0.4$ for (\textit{a}, \textit{c}) \emph{Group-Re} and (\textit{b}, \textit{d}) \emph{Group-ks} cases. The top row is normalised by local $U_{\tau 2}$ (inner scaling) and the bottom row is normalised by the rough-wall $U_{\tau 0}$ and boundary layer thickness $\delta_{99}$ (outer scaling). The solid lines are the smooth-wall reference \citep{marusic2015evolution, squire2016comparison} interpolated to $Re_{\tau}$ matched to the corresponding rough-to-smooth profile. The colour of the symbols and lines indicates $Re_{\tau 0}$ and $k_{s0}^+$ in (\textit{a}, \textit{c}) and (\textit{b}, \textit{d}), respectively, as defined in table \ref{tab:cases}.} 
\label{fig:uvar_Re_ks}
\end{figure}

Finally, the effect of $Re_{\tau 0}$ and $k_{s0}^+$ on the strength of outer-layer structures are examined by plotting the inner-scaled variance profiles of \emph{Group-Re} and \emph{Group-ks} cases at $\hat{x}/\delta_0\approx0.4$ in figure \ref{fig:uvar_Re_ks}(\textit{a}) and (\textit{b}). Note that the difference in $Re_{\tau}$ between \emph{Group-ks} cases is due to the variation in $U_{\tau2}/U_{\tau 0}$, the ratio of friction velocities across the step change. The variance becomes higher with the increase of $Re_{\tau 0}$ and $k_{s0}^+$. However, within the parameter range in this study, the dependence of the inner-scaled variance on $Re_{\tau 0}$ is much less than that on $k_{s0}^+$. This observation will be revisited in the context of energy spectra in the following section. A collapse in the outer layer is observed in figure 21(\textit{c}) and (\textit{d}), on the other hand, when the outer length scale $\delta_{99}$ and the rough-wall friction velicity $U_{\tau 0}$ are used to normalise the profiles, indicating that the outer part of the recovering boundary layer continues to obey outer-layer similarity based on upstream (rough-wall) conditions.

\section{Premultiplied energy spectrum} \label{sec:big_spectrum}

In this section, we examine the effect of $Re_{\tau 0}$ and $k_{s0}^+$ on the turbulent energy distribution at various wavelengths (i.e. the premultiplied energy spectrum) of flows downstream of the rough-to-smooth transition, especially in light of the large-scale influence in the near-wall region noted in \S\ref{sec:big_turb} and in \cite{MogengJFM2019}.

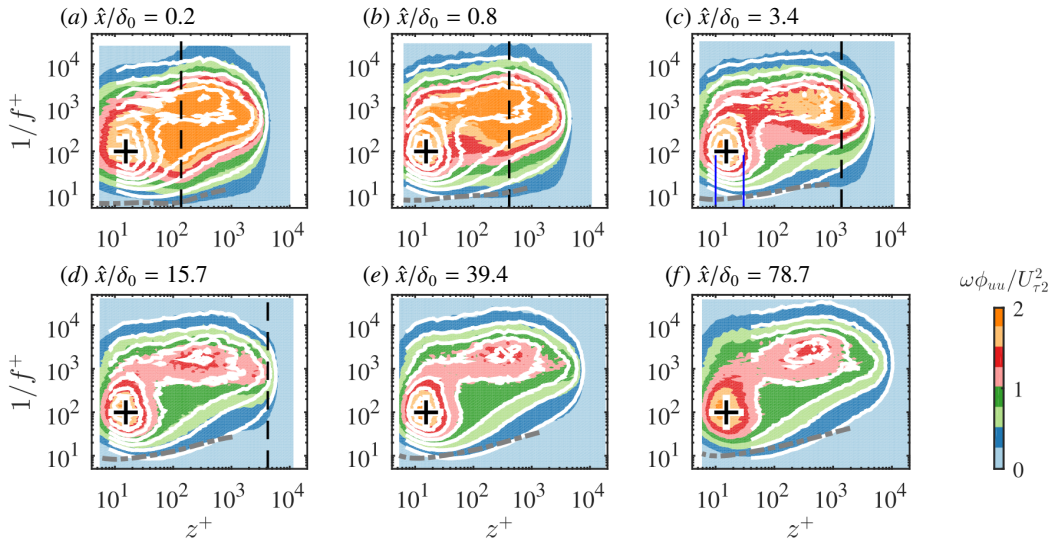
\begin{figure}
    \centering
\setlength{\unitlength}{1cm}
\begin{picture}(14, 7.1)
\put(0,0){\includegraphics[scale = 0.92]{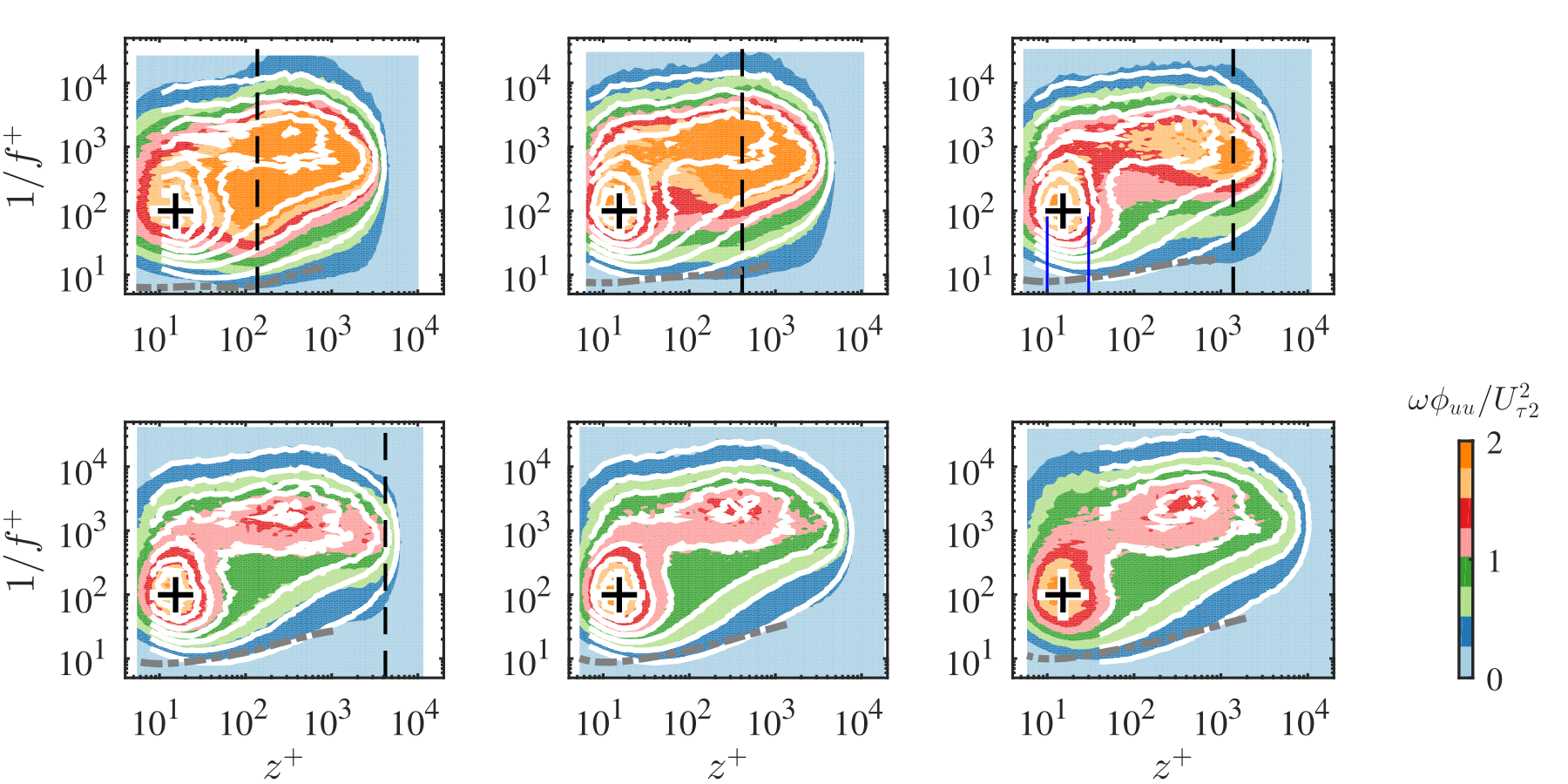}}
\put(0.7,6.8){(\textit{a}) $\hat{x}/\delta_0 = 0.2$} 
\put(4.7,6.8){(\textit{b}) $\hat{x}/\delta_0 = 0.8$}
\put(8.7,6.8){(\textit{c}) $\hat{x}/\delta_0 = 3.4$}
\put(0.7,3.4){(\textit{d}) $\hat{x}/\delta_0 = 15.7$} 
\put(4.7,3.4){(\textit{e}) $\hat{x}/\delta_0 = 39.4$}
\put(8.7,3.4){(\textit{f}) $\hat{x}/\delta_0 = 78.7$}
\end{picture}
    \caption{Viscous-scaled premultiplied energy spectrum $\omega \phi_{uu}/U^2_{\tau 2}$ as a function of increasing fetch $\hat{x}/\delta_0$ from the rough-to-smooth transition. From (\textit{a}) to (\textit{f}), $\hat{x}/\delta_0 = 0.2,\,0.8,\,3.4,\,15.7,\,39.4$ and 78.7. The colour contours correspond to the rough-to-smooth case \texttt{Re07ks16}, and the white contour lines are interpolated from a reference smooth-wall experimental dataset to matched $Re_{\tau}$. Contour levels are chosen at $\omega \phi_{uu}/U^2_{\tau 2}$ = 0 to 2 with an increment of 0.25. The vertical black dashed line represents the location of $\delta_i$. The black plus is at $z^+=15$ and $1/f^+=95$ (equivalent to $\lambda_x^+\approx 1000$). The thick grey dot-dashed lines represent $1/f^+_{\eta}$, the Kolmogorov time scale. The blue box in (\textit{c}) is at $5<1/f^+<90$ and $10<z^+<30$, showing the region where the spectrum fit is performed. }
    \label{fig:Spectrum_recovery}
\end{figure}

The premultiplied energy spectrum $\omega \phi_{uu}/U^2_{\tau 2}$ of case \texttt{Re07ks16} is shown in figure \ref{fig:Spectrum_recovery}, where $\omega = 2\pi f$ is the angular frequency, $f$ is the frequency (corresponding to the wavenumber in the spatial domain), $\phi_{uu}$ is the energy spectrum of the streamwise velocity fluctuation ($\int_0^{\infty} \phi_{uu} \mathrm{d} \omega = \overline{u^2}$), and $U_{\tau 2}$ is the friction velocity measured from the OFI experiments. The spectrograms are computed from hotwire time-series data. Since the flow is heterogeneous in $x$, we refrain from converting the spectrum from temporal to the spatial domain, which also side-steps uncertainties in convection velocity for rough-wall flows \citep{squire2017applicability}. The coloured contours are the current rough-to-smooth data, and the white contour lines are interpolated from a reference smooth-wall experimental dataset \citep{marusic2015evolution,squire2016comparison} at matched $Re_{\tau}$, which ensures that the energy diminishes to zero at the same wall-normal height in viscous units in both the rough-to-smooth case and the smooth-wall reference (matched $\delta_{99}^+$). The coloured contour quite closely follows the white contour lines in the near-wall, high-frequency region of the spectrum, while the difference becomes more noticeable at lower frequencies and with increasing distance away from the wall. Figure \ref{fig:Spectrum_recovery_diff} shows the difference between the rough-to-smooth spectrum and the reference smooth-wall spectrum, defined as
\begin{figure}
    \centering
\setlength{\unitlength}{1cm}
\begin{picture}(14,7.1)
\put(0,0){\includegraphics[scale = 0.92]{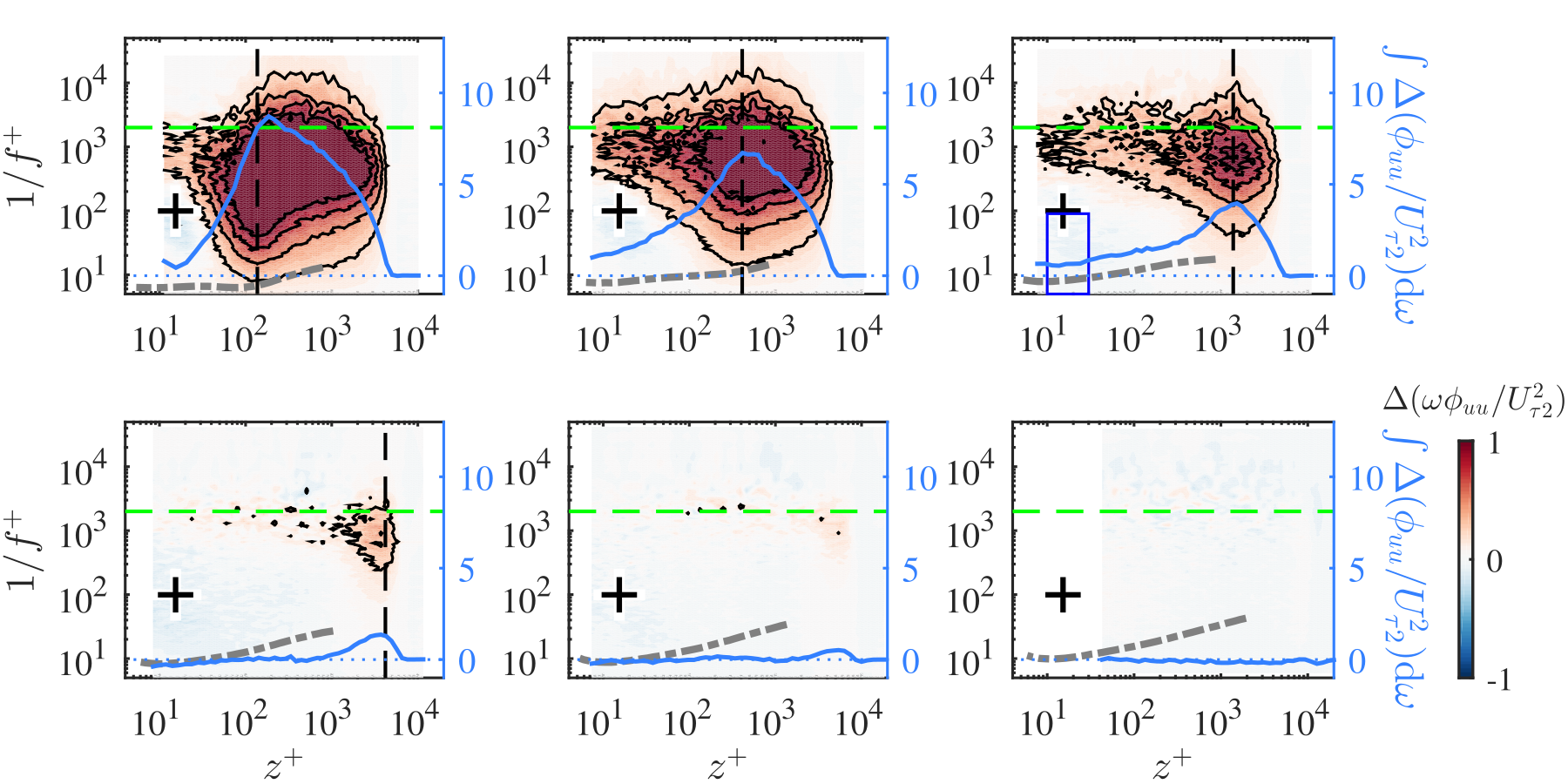}}
\put(0.7,6.8){(\textit{a}) $\hat{x}/\delta_0 = 0.2$} 
\put(4.7,6.8){(\textit{b}) $\hat{x}/\delta_0 = 0.8$}
\put(8.7,6.8){(\textit{c}) $\hat{x}/\delta_0 = 3.4$}
\put(0.7,3.4){(\textit{d}) $\hat{x}/\delta_0 = 15.7$} 
\put(4.7,3.4){(\textit{e}) $\hat{x}/\delta_0 = 39.4$}
\put(8.7,3.4){(\textit{f}) $\hat{x}/\delta_0 = 78.7$}
\end{picture}
    \caption{The difference between the viscous-scaled premultiplied spectrum of \texttt{Re07ks16} and the smooth-wall reference (matched $Re_{\tau}$) at streamwise locations corresponding to figure \ref{fig:Spectrum_recovery}.  The four black contour lines indicate  $\Updelta(\omega \phi_{uu}/U^2_{\tau 2}) = $ 0.25, 0.5, 0.75 and 1. The vertical dashed black line represents the location of $\delta_i$ and the horizontal green dashed line is at $1/f^+ = 2\times10^3$. The black plus is at $z^+=15$ and $1/f^+=95$ (equivalent to $\lambda_x^+\approx 1000$). The thick grey dot-dashed lines represent $1/f^+_{\eta}$, the Kolmogorov time scale. The blue box in (\textit{c}) is at $5<1/f^+<90$ and $10<z^+<30$, showing the region where the spectrum fit is performed. The blue line is the difference in $\phi_{uu}/U_{\tau 2}^2$ integrated across all wavelengths. }
    \label{fig:Spectrum_recovery_diff}
\end{figure}
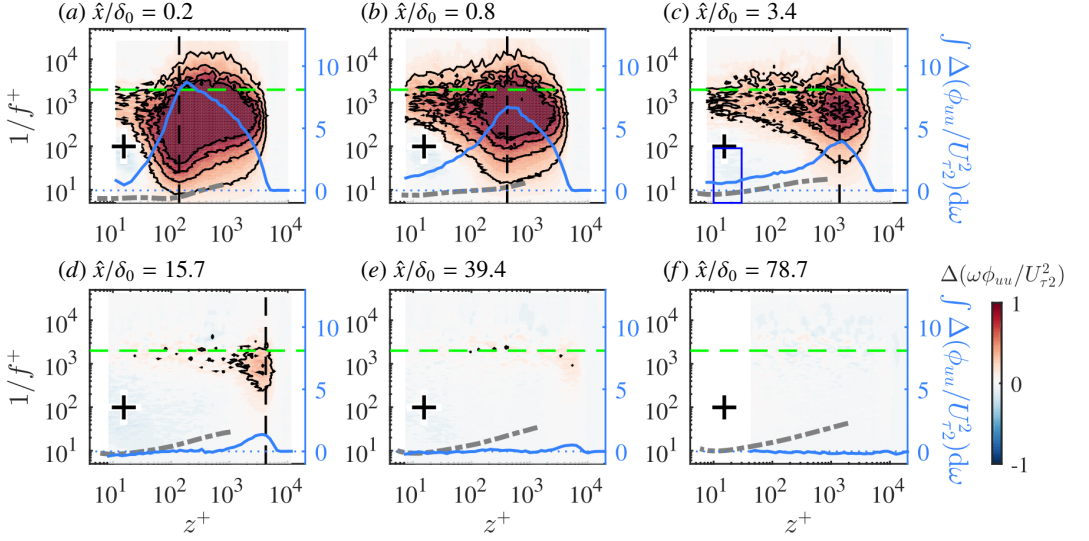
\begin{equation}
\Updelta (\omega\phi_{uu}/U_{\tau 2}^2) \equiv (\omega\phi_{uu}/U_{\tau 2}^2)_{R\rightarrow S} - (\omega\phi_{uu}/U_\tau^2)_{S}
\end{equation}
for the downstream locations from $\hat{x}/\delta_0 = 0.2$ (figure \ref{fig:Spectrum_recovery_diff}\textit{a}) to 78.7 (figure \ref{fig:Spectrum_recovery_diff}\textit{f}). Similar to the previous observation \citep{ismail2018simulations, MogengJFM2019}, the good agreement in the near-wall, small-scale energy spectrum between the rough-to-smooth case and the smooth-wall reference persists at higher $Re_{\tau}$ values in the present study. We will refer to this near-wall, high-frequency region where $\Updelta (\omega\phi_{uu}/U_{\tau 2}^2) $ has reached 0 as `the fully-recovered region'. Structures residing above the IBL, which scale on the rough-wall friction velocity $U_{\tau 0}$, are over-energised compared to the smaller local $U_{\tau 2}$ (the IBL location is marked by the vertical black dashed line in figures  \ref{fig:Spectrum_recovery} and \ref{fig:Spectrum_recovery_diff}). As depicted in figures \ref{fig:Spectrum_recovery_diff}(\textit{a-c}), these structures leave a `footprint' of excess energy at $1/f^+\sim\mathcal{O}(10^3)$ in the near-wall region, while the energy distribution at smaller scales is only slightly modified. 

\begin{figure}
    \centering
\setlength{\unitlength}{1cm}
\begin{picture}(15.24,9.65)
\put(0,0){\includegraphics[scale = 0.91]{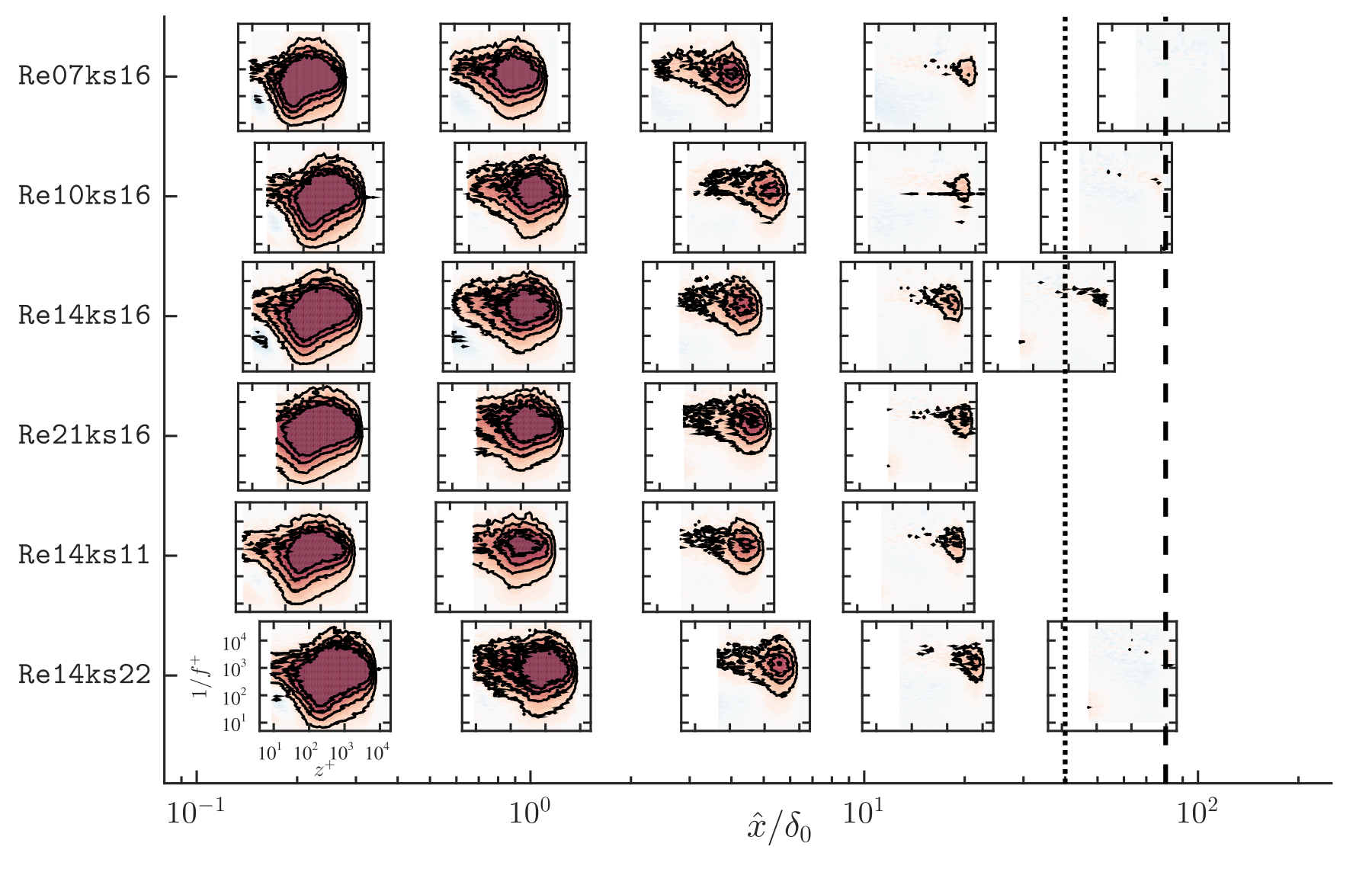}}
\end{picture}
\caption{The excess energy spectrum of all 6 cases at various representative streamwise locations. The $x$-coordinate of the centre of each plot represents the streamwise location where the measurement was obtained. The colour bar is the same as in figure \ref{fig:Spectrum_recovery_diff_xhat}. The vertical dotted and dashed lines are at $\hat{x}/\delta_0 = 40$ and 80, respectively.}
\label{fig:Spectrum_recovery_diff_map}
\end{figure}

\begin{figure}
    \centering
\setlength{\unitlength}{1cm}
\begin{picture}(14,6.5)
\put(0,0){\includegraphics[scale = 0.92]{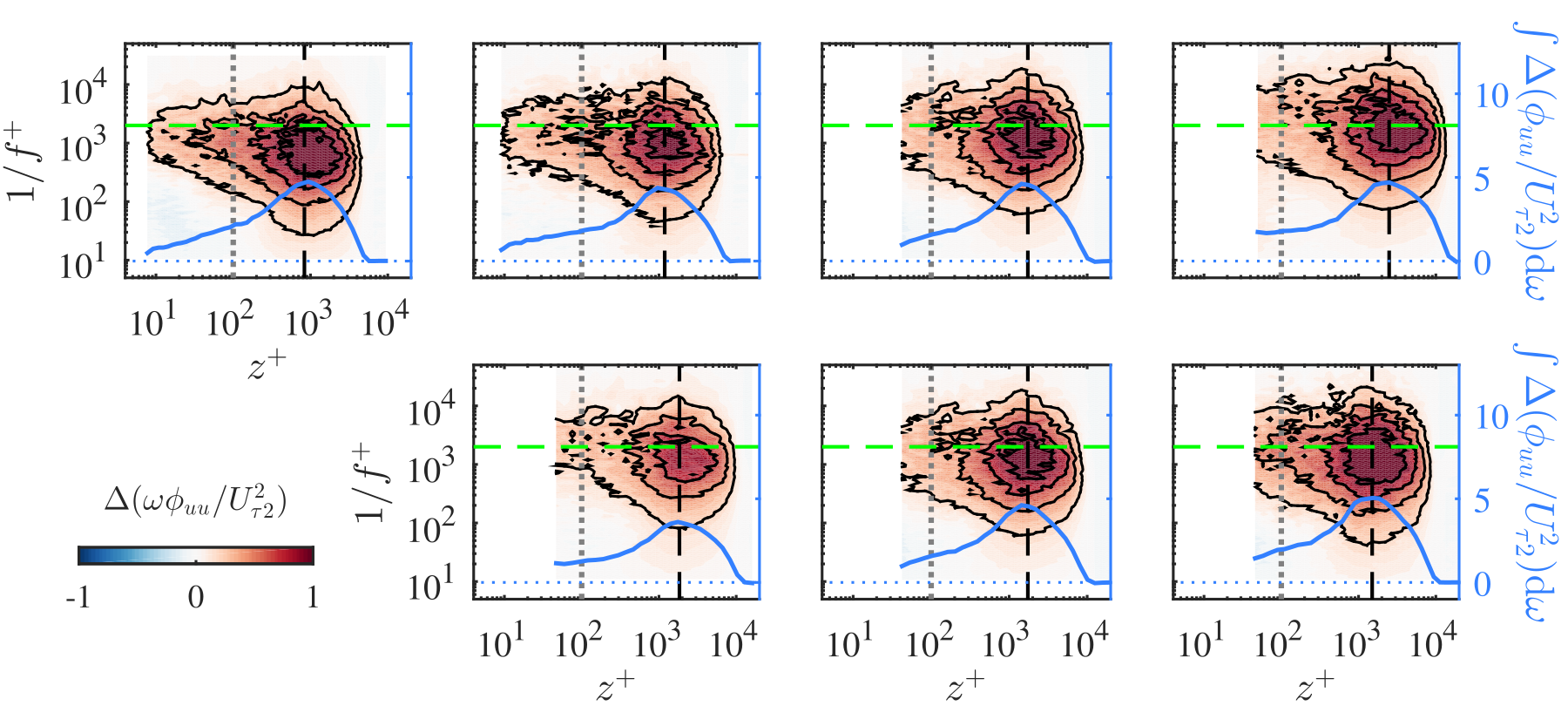}}
\put(0.8,6.2){(\textit{a}) \texttt{Re07ks16}} 
\put(3.9,6.2){(\textit{b}) \texttt{Re10ks16}}
\put(7.0,6.2){(\textit{c}) \texttt{Re14ks16}}
\put(10.1,6.2){(\textit{d}) \texttt{Re21ks16}}
\put(3.9,3.3){(\textit{e}) \texttt{Re14ks11}}
\put(7.0,3.3){(\textit{f}) \texttt{Re14ks16}}
\put(10.1,3.3){(\textit{g}) \texttt{Re14ks22}}
\end{picture}
    \caption{The difference between the viscous-scaled premultiplied spectrum and the smooth-wall reference (matched $Re_{\tau}$) at $\hat{x}/\delta_0=2.0$. The first row (\textit{a-d}) and the second row (\textit{e-g}) are \emph{Group-Re} and \emph{Group-ks} cases, respectively. $Re_{\tau 0}$ increases from left to right in the first two rows, and $k_{s0}^+$ increases from left to right in the last row. In each plot, the four black contour lines indicate  $\Updelta(\omega \phi_{uu}/U^2_{\tau 2}) = $ 0.25, 0.5, 0.75 and 1. The vertical dashed black line represents the location of $\delta_i$. The vertical grey dotted line shows $z^+ = 100$. The horizontal green dashed line is at $1/f^+ = 2\times10^3$, serving as a reference of the large-scale footprint locations. The blue curve is the difference in $\phi_{uu}/U_{\tau 2}^2$ integrated across all wavelengths. }
    \label{fig:Spectrum_recovery_diff_xhat}
\end{figure}

The fully-recovered region encapsulates turbulence scales from those associated with the near-wall cycle (as marked by the plus symbol in figure \ref{fig:Spectrum_recovery}\textit{a}) down to the dissipation scales, which can be characterised by the Kolmogorov time scale\citep{Tennekes1972,Pope2000}
\begin{equation}
1/f^+_{\eta} = (\nu/\epsilon)^{1/2},
\end{equation}
where the dissipation rate $\epsilon$ is estimated from the streamwise energy spectrum as
\begin{equation}
\epsilon = 15\nu\int_0^{\infty}\frac{\omega^2}{U^2}\phi_{uu}\mathrm{d}\frac{\omega}{U}.
\end{equation}
$1/f^+_{\eta}$ is shown in figures \ref{fig:Spectrum_recovery} and \ref{fig:Spectrum_recovery_diff} by the thick grey dot-dashed line. The wall-normal extent of the fully-recovered region is initially limited to the buffer region ($10<z^+<30$) at $\hat{x}/\delta_0=0.2$, and gradually expands further away from the wall with the growth of the IBL. 

Most previous laboratory measurements only cover a downstream fetch of approximately $20 \delta_0$ or less, where usually, the flow still has not reached a full recovery to the new wall condition. In the present study, we are able to measure up to $120\delta_0$ for case \texttt{Re07ks16}, which enables us to study the recovery of the flow in the far field. A complete recovery of the energy spectrum is achieved at $\hat{x}/\delta_0 = 78.7$, as shown in figure \ref{fig:Spectrum_recovery}(\textit{f}) and figure \ref{fig:Spectrum_recovery_diff}(\textit{f}), from which we can estimate that a complete recovery of the energy spectrum can be expected somewhere in the range from $39.4\delta_0$ to $78.7\delta_0$ downstream of the roughness transition. We lack an intermediate measurement between these locations due to the logarithmic spacing of streamwise measurement stations. We further examine whether the recovery length remains the same in other cases with different $Re_{\tau 0}$ or $k_{s0}^+$ by compiling the excess energy spectrum contours at several representative streamwise locations into one figure (figure \ref{fig:Spectrum_recovery_diff_map}), with the $x$-coordinate of the centre of each panel as the measurement location. At $\hat{x}/\delta_0\approx15$, all cases have very similar contours of excess energy. The measurement locations become sparse downstream of this point, and there are no measurements beyond this location for \texttt{Re21ks16} and \texttt{Re14ks11}, the two cases with the longest sandpaper patch. Some excess energy remaining close to the edge of the boundary layer can be observed in \texttt{Re10ks16}, \texttt{Re14ks22} and \texttt{Re14ks16} at $\hat{x}/\delta_0\approx 40$ (vertical dotted line), while a complete recovery with no distinguishable excess energy is first observed at $\hat{x}/\delta_0\approx80$ (vertical dashed line) in \texttt{Re07ks16}. In summary, from the limited range of $Re_{\tau0}$ and $k_{s0}^+$ investigated here, no discernible dependence on $Re_{\tau 0}$ or $k_{s0}^+$ in the recovery trends of the excess energy spectrum can be concluded.

Regardless, the noteworthy feature from these spectra is that it takes a longer fetch downstream for the energy spectrum to relax completely to the smooth-wall state than for the IBL to outgrow the original boundary layer, which is approximately $26.5\delta_0$ as predicted by extrapolating the power-law fit (\ref{Eq:IBL}). Comparing figure \ref{fig:Cf} and figure \ref{fig:Spectrum_recovery_diff}, it appears that, within experimental uncertainty, $C_f$ achieves a complete recovery from the roughness transition in a shorter fetch ($20\delta_0$) compared to the energy spectrum ($40\delta_0-80\delta_0$). Qualitatively similar observations have been reported by  \cite{rouhi2018}, \cite{ismail2018simulations} and \cite{sridhar2018} in their numerical studies. The streamwise location where $C_f$ has reached the complete recovery seems to approximately coincide with the location where the energetic large-scale footprint in the near-wall region vanishes: at $\hat{x}/\delta_0 = 15.7$ (figure \ref{fig:Spectrum_recovery_diff}\textit{d}), this footprint is already becoming very weak for $z^+ < 100$ (marked by the horizontal dashed lines in figure \ref{fig:Spectrum_recovery_diff}).

We further examine the effect of $Re_{\tau 0}$ and $k_{s0}^+$ on the recovery behaviour of the energy spectrum utilising both \emph{Group-Re} and \emph{Group-ks}. The energy spectra are interpolated to a matched streamwise fetch $\hat{x}/\delta_0 = 2$, and summarised in figure \ref{fig:Spectrum_recovery_diff_xhat}. We begin by analysing the less complicated \emph{Group-ks} cases, as shown in figure \ref{fig:Spectrum_recovery_diff_xhat}(\textit{e-g}). $\delta_i$ is almost linearly proportional to $\hat{x}$ according to equation (\ref{Eq:IBL}), which implies that both $\delta_i U_{\tau 0}/\nu$ and $\delta_i/{\delta_{99}}$ are also approximately matched in these \emph{Group-ks} cases at the matched streamwise location. The excess energy above the IBL is centred around the same location in the frequency domain (similar $1/f^+$ values), leaving a footprint in the near-wall region as marked by the dashed green lines at $1/f^+ = 2000$. In this study, a higher $k_{s0}^+$ leads to a higher $U_{\tau 0}$ to local $U_{\tau 2}$ ratio, so the $U_{\tau 0}$-scaled rough-wall structures in the outer layer are more energetic compared to the $U_{\tau 2}$-scaled small-scale energy near the wall. Therefore, it is expected that under $U_{\tau 2}$ scaling the magnitude of both the outer large-scale motions and their footprints increases with increasing $k_{s0}^+$, as evidenced by the increasing intensity at a wall-normal location fixed in viscous units (marked by the grey line at $z^+ = 100$) from (\textit{e}) to (\textit{g}). 

For \emph{Group-Re} cases at matched $\hat{x}/\delta_0 = 2$ (figure \ref{fig:Spectrum_recovery_diff_xhat}\textit{a-d}), an increase in $\delta_i^+$ is expected with an increasing $Re_{\tau 0}$, although $\delta_i/\delta_{99}$ remains almost constant. With a near constant friction velocity ratio $U_{\tau 0}/U_{\tau 2}$, this leads to an almost constant excess energy level in the outer large-scale motions, as evidenced by figure \ref{fig:Spectrum_recovery_diff_xhat}(\textit{a-d}) and also supported by the outer-layer similarity hypothesis above the IBL. As $Re_{\tau 0}$ increases, the near-wall footprint is observed to shift to a larger $1/f^+$ from (\textit{a}) to (\textit{d}), while its magnitude decreases slightly. We speculate that the footprinting effect felt in the near-wall region is limited by the increased spatial separation as the large-scale motions are farther away from the wall in viscous units at a greater $Re_{\tau 0}$. 

To conclude from these observations, the viscous-scaled time period $1/f^+$ of the near-wall footprint from the over-energised large-scale motions increases with $Re_{\tau0}$ while it is little affected by $k_{s0}^+$. The excess energy in the footprint is found to increase when the large-scale structures in the outer layer are more energised (due to high $k_{s0}^+$) or closer to the wall in viscous units (due to lower $\delta_i^+$ which occurs for lower $Re_{\tau 0}$ at fixed $\hat{x}/\delta_0$).

\subsection{Estimating $U_{\tau 2}$ from the premultiplied energy spectrum}
Based on the observation that the energy distribution at the high-frequency end in the near-wall region recovers to the smooth-wall equilibrium state faster than the lower frequencies, \cite{MogengJFM2019} previously suggested an alternative method to extract $U_{\tau 2}$ from the premultiplied energy spectrum when a direct wall-shear stress measurement such as OFI is not available. The surrogate method was tested on only one set of data at $Re_{\tau 0}=4100$. Here we further examine the validity of this method over a wider range of $Re_{\tau 0}$ and $k_{s0}^+$.
\begin{figure}
\centering
\setlength{\unitlength}{1cm}
\begin{picture}(13,10.2)
\put(0,5.1){\includegraphics[scale = 0.95]{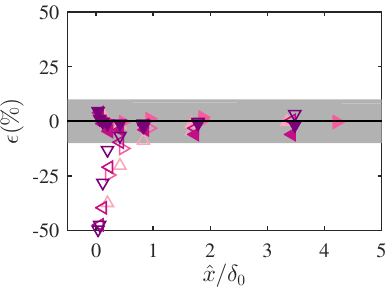}}
\put(6.5,5.1){\includegraphics[scale = 0.95]{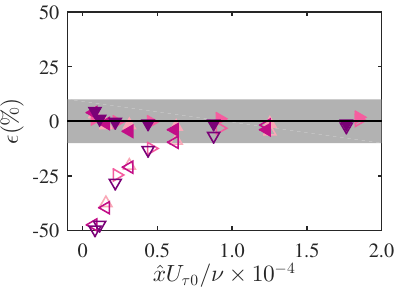}}
\put(0.1,9.9){(\textit{a})}
\put(6.6,9.9){(\textit{b})}
\put(0,0){\includegraphics[scale = 0.95]{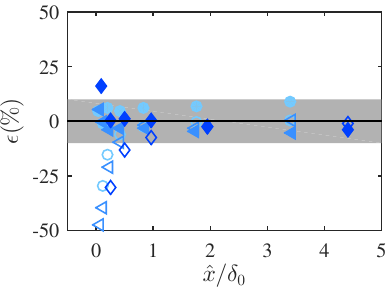}}
\put(6.5,0){\includegraphics[scale = 0.95]{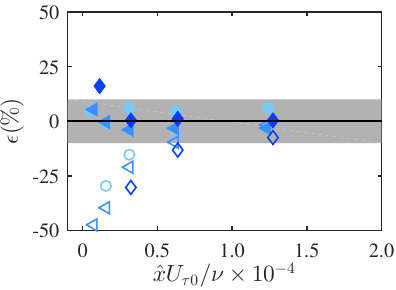}}
\put(0.1,4.8){(\textit{c})}
\put(6.6,4.8){(\textit{d})}
\end{picture}
\caption{$\epsilon$, the error in the estimated $C_{f}$ relative to the OFI results, as defined in (\ref{eq:epsilon_utau_surro}). Open symbols represent a buffer region fit and solid symbols are for the spectrum fit. The shaded band covers $-10\%$ to $10\%$ on the vertical axis. The first row is for \emph{Group-Re} cases, and the second row is for \emph{Group-ks} cases. The colour of the symbols indicates $Re_{\tau 0}$ and $k_{s0}^+$ in the first and second row, respectively, as defined in table \ref{tab:cases}. The horizontal axis is the fetch normalised by the outer length scale $\delta_0$ in (\textit{a,c}) and the viscous scale in (\textit{b,d}).}
\label{fig:utau_surro_big}
\end{figure}
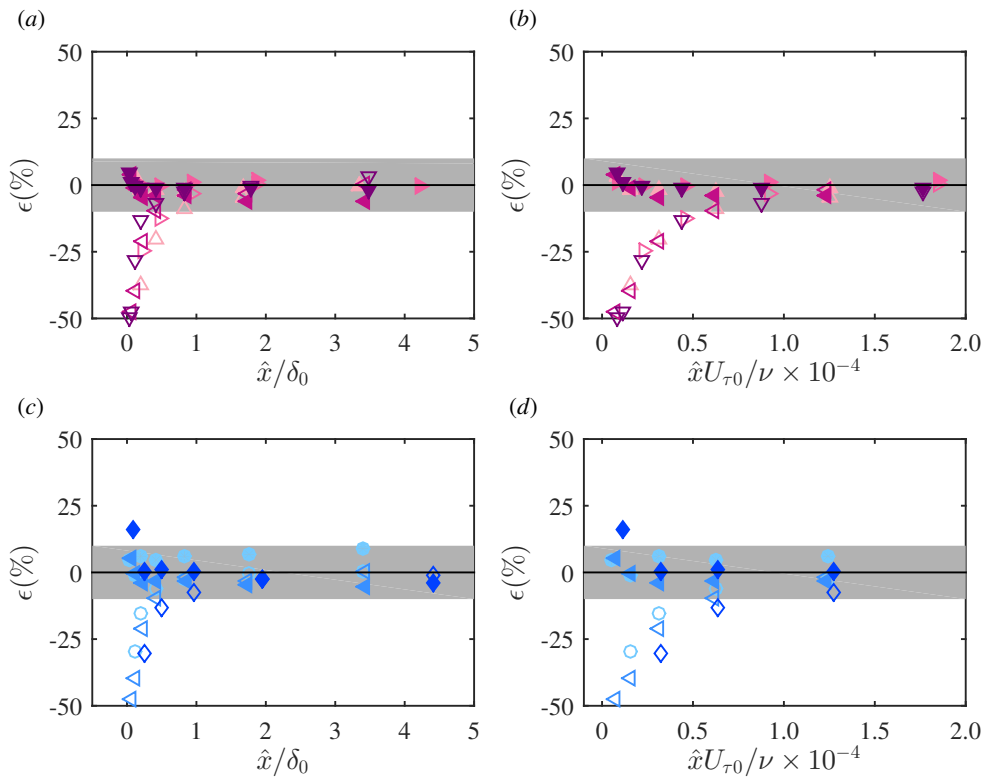

We first provide a brief summary of the method described previously. A rectangular region in the spectrogram plot bounded by the limits $10<z^+<30$ and $5<1/f^+<90$ is chosen, and the difference between the viscous scaled energy spectra for the rough-to-smooth and the reference smooth case $\Updelta (\omega \phi_{uu}/U_{\tau 2}^2)$ is then minimised across this region by varying $U_{\tau 2}$. \cite{MogengJFM2019} compared the error in the estimation of the spectrum fit method with that from a buffer-region fit of the mean velocity profile, representing conventional methods that assume a canonical or equilibrium mean velocity profile. The proposed spectrum-based technique exhibited much better performance, especially immediately downstream of the roughness transition. \cite{MogengJFM2019} defined $\epsilon$ as the relative error between the skin-friction coefficient determined from a particular indirect technique and the OFI results as
\begin{equation}
\epsilon \equiv \frac{C_{f}|_{M}-C_{f}|_{OFI}}{C_{f}|_{OFI}}\times100\%.
\label{eq:epsilon_utau_surro}
\end{equation}
The subscript $M$ stands for the method, which can be either the spectrum fit or the buffer-region fit. \cite{MogengJFM2019} found that the error of the spectrum fit falls between $\pm10\%$ at all streamwise locations, while an underestimation of over $25\%$ is observed for the buffer-region fit close to the roughness transition. 

In this study we test the efficacy of the surrogate spectral method of estimating $U_{\tau2}$ over a much wider range of $Re_{\tau 0}$ in figure \ref{fig:utau_surro_big}. As shown in (\textit{a,b}) which compares \emph{Group-Re} results, the spectrum fit method retains a good agreement with the reference OFI data for all cases, while the buffer-region fit produces an underestimation close to the roughness transition \citep[as noted by][]{MogengJFM2019}. The spectrum fit method also exhibits similar performance for \emph{Group-ks} cases in  (\textit{c,d}). We have previously reported a minimum fetch of $5\delta_0$ after the roughness transition before the buffer-region method can be safely used. An even longer fetch ($\hat{x}\gtrsim8h$) is required for the open-channel DNS dataset (which has a much lower $Re_{\tau0}$). For the cases shown in figure \ref{fig:utau_surro_big}(\textit{a}), the error in the buffer-region fit diminishes in a fetch that is typically smaller than $2\delta_0$, and the required fetch for reduced error decreases even further as $Re_{\tau 0}$ increases. The results indicate that the minimum fetch scales in viscous units rather than $\delta_0$ or $h$, hence we present $\epsilon$ versus $\hat{x} U_{\tau 1}/\nu$ in figure \ref{fig:utau_surro_big}(\textit{b}). The results exhibit a much better collapse for the different Reynolds number cases than in figure \ref{fig:utau_surro_big}(\textit{a}) and we observe that a nominally zero $\epsilon$ of the buffer-region fit method is observed at $\hat{x}U_{\tau 0}/\nu \gtrsim 1.0\times10^4$. The error in the buffer fit method is also seen to increase with increasing $k_{s0}^+$ in figure \ref{fig:utau_surro_big}(\textit{d}), because the boundary layer is recovering from a greater change. 

We may conclude that, for a certain $k_{s0}^+$ (and possibly a certain step height), the minimum fetch required before a buffer-region fit can give good wall-shear stress estimates is approximately constant in upstream viscous units. For the buffer-region fit method to be valid, it essentially requires a full recovery of the mean velocity profile within the buffer layer, i.e. the edge of the equilibrium layer should exceed the upper limit of the fitting region. As the upper limit of the buffer region is usually fixed in viscous units, it is expected that the minimum fetch for the equilibrium layer to reach such a height will also be a constant when scaled by viscous units. In fact, the recovery length of the buffer-region fit method can be estimated from the blending function as discussed in \S\ref{sec:blending}. Using the rough approximation $\delta_i=0.1\hat{x}$ and a threshold of $E=0.95$, the viscous-scaled fetch required for $\delta_e^+$ to reach the upper limit of the buffer region (i.e. 100 wall units) is found to be $\hat{x}U_{\tau 0}/\nu \approx 1.0\times10^4$. Consequently, we can also speculate that a log-region fit to estimate $U_{\tau 2}$ (classic Clauser) would require that the equilibrium layer should reach $0.15\delta_{99}$, and hence would require a much longer fetch that would scale in $\delta_0$.

\section{Conclusions} \label{sec:conclusions}
In this study, we present an experimental dataset documenting the evolution of a turbulent boundary layer downstream of a rough-to-smooth surface transition. To redress previously reported uncertainties with $C_f$ recovery, the skin friction downstream of the transition is measured directly using OFI technique. In an attempt to unpack the effects of $k_{s0}^+$ and $Re_{\tau 0}$ (which often both vary between studies), the experimental cases are classified into two groups based on their flow conditions. For all \emph{Group-Re} cases, a nominally constant $k_{s0}^+\approx160$ is achieved with $Re_{\tau 0}$ ranging from 7100 to 21000, while a nominally constant $Re_{\tau 0}\approx 14000$ is achieved for all \emph{Group-ks} cases with $k_{s0}^+$ ranging from 111 to 228. Our main findings are summarised as below.

\begin{itemize} 
\item The recovery of $C_f$ follows a single trend with $\hat{x}/\delta_0$ when normalised by $C_{fe}$ (the skin-friction coefficient expected at the same Reynolds number using the smooth-wall relationship), showing little $Re_{\tau 0}$ or $k_{s0}^+$ dependence (figure \ref{fig:Cf}).

\item A method to extract $\delta_i$, the IBL height, from the turbulence intensity profile is described. $\delta_i$ values obtained using this method are compared with the results following the definition of \cite{Antonia1971a} (which is a commonly used method in the literature), and we show that the present definition leads to a slightly higher $\delta_i$ as it tends to pick up the upper limit of the IBL. $\delta_i$ computed using both definitions follows a power-law growth trend with increasing development downstream of the transition with an exponent close to 0.8. Little $Re_{\tau 0}$ or $k_{s0}^+$ dependence is observed in $\delta_i$, and $\delta_i$ is expected to outgrow the local $\delta_{99}$ at $\hat{x} = 26.5\delta_0$ according to the power law (figure \ref{fig:IBL}).

\item The mean velocity profile downstream of a rough-to-smooth change is found to undershoot the canonical smooth-wall profile, and it can be modelled by blending the corresponding rough-wall and smooth-wall profiles through a blending function $E(z)$. This function $E(z)$ can be approximated by an error function with empirically determined parameters, and when $\delta_i$ is selected as the length scale, it also exhibits self-similarity with no dependence on $Re_{\tau 0}$ or $k_{s0}^+$ observed within the range of parameters investigated (figure \ref{fig:Blend_Gamma}).

\item A complete recovery of the energy spectrum occurs within a fetch between $40\text{--}80\delta_0$ for case \texttt{Re07ks16}, longer than the fetch required for $\delta_i$ to outgrow $\delta_{99}$ as predicted by the power-law relation. The recovery length has shown little dependence on $Re_{\tau 0}$ or $k_{s0}^+$ when normalised by $\delta_0$ (figure \ref{fig:Spectrum_recovery_diff}).  

\item The over-energised inner peak observed in the streamwise turbulence intensity profiles is found to be contributed by the large-scale motion footprints in the near-wall region. The viscous-scaled time period ($1/f^+$) of these footprints increases with $Re_{\tau0}$ while it is little affected by $k_{s0}^+$. The excess energy in the footprint is found to increase when the large-scale structures in the outer layer are more energised (due to high $k_{s0}^+$) or closer to the wall in viscous units (due to lower $\delta_i^+$ which occurs for lower $Re_{\tau 0}$ at fixed $\hat{x}/\delta_0$) (figure \ref{fig:Spectrum_recovery_diff_xhat}). 

\item The spectrum fit method to extract local friction velocity proposed by \cite{MogengJFM2019} exhibits a good performance over a wide range of $Re_{\tau 0}$. For an equilibrium layer with a growth rate similar to the present study, the error in the estimation from a buffer-region fit of the mean velocity profile is considered to diminish after $\hat{x}U_{\tau 0}/\nu\gtrsim1.0\times10^4$ (figure \ref{fig:utau_surro_big}).

\item The length and velocity scales in the outer layer remain relatively unchanged right after the rough-to-smooth change. Farther downstream, both $C_{f,\mathrm{out}}$ and $\delta_{99}$ begin to slowly adapt to the smooth-wall state, presumably as the over-energised large-scale structures in the outer layer decay and are replaced by events that reflect the new surface condition. A noticeable deviation from the rough-wall trend only appears after approximately $10\delta_0$ downstream of the transition (figures \ref{fig:delta99_out} and \ref{fig:CfoU_Uvar}).

\end{itemize}

To conclude, downstream of a rough-to-smooth change, many statistics of the recovering flow, such as $C_f$, $\delta_i$ and $E(z)$, are little affected by $Re_{\tau 0}$ and $k_{s0}^+$ when scaled properly. This observation is meaningful in the application of high Reynolds number atmospheric or industrial flows beyond laboratory ranges. In the current study, the range of $k_{s0}^+$ is limited compared to other studies such as \cite{Hanson2016}. Hence, a wider range of $k_{s0}^+$ values need to be investigated in future works in search of potential $k_{s0}^+$ effects.

\section*{Acknowledgements}
This research was partially supported under the Australian Research Council's Discovery Projects funding scheme (project DP160103619).

\section*{Declaration of Interests}
 The authors report no conflict of interest.

\appendix

\section{Evolution of the outer layer}\label{sec:big_outer}
Traditionally, studies on the flow downstream of a roughness transition have tended to focus on the thickness of the developing IBL and the local friction velocity. The length and velocity scales in the outer layer are usually assumed to remain constant and the evolution of the outer layer in the streamwise direction is often neglected \citep[for example][]{Elliott1958, panofsky1964change, Chamorro2009}. In addition, the mean velocity profile in the outer layer is often characterised by a logarithmic law only and the boundary layer thickness $\delta_{99}$ often does not feature in the modelled or assumed profiles. These assumptions may be adequate in a short fetch downstream of the transition, where the recovery of the flow is rapid and the Reynolds number does not change substantially. However, even in a turbulent boundary layer developing on a homogeneous surface in a zero-pressure gradient, the streamwise evolution of the flow leads to an increase in the boundary layer thickness and a decrease in the friction velocity with increasing $x$. The decrease in the outer-layer friction velocity scale $U_{\tau,\mathrm{out}}$ has been shown to be enhanced following a rough-to-smooth change \citep{Hanson2016}. Here, utilising the current experimental dataset, the far-field evolution of $\delta_{99}$ and $U_{\tau, \mathrm{out}}$ is investigated and compared to that of a homogeneous rough-wall or smooth-wall boundary layer.

\begin{figure}
\centering
\setlength{\unitlength}{1cm}
\begin{picture}(6.5,5.1)
\put(0,0){\includegraphics[scale = 0.95]{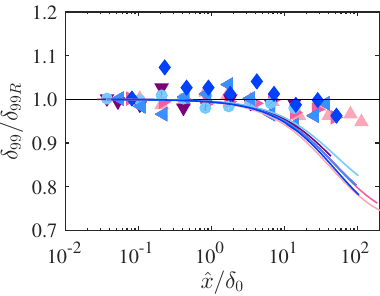}}
\end{picture}
\caption{$\delta_{99}$ determined from the measured mean velocity profiles downstream of the rough-to-smooth normalised by $\delta_{99R}$, the boundary layer thickness for a homogeneously rough surface. The solid black line is at $\delta_{99}/\delta_{99R} = 1$. Reference curves are included for comparison, which are predicted using the von K\'arm\'an momentum integral equation assuming that the boundary layer immediately adjusts to a smooth-wall profile with matched $\delta_{99}$ right after the rough-to-smooth transition and shown by solid line with a colour corresponding to each case.}
\label{fig:delta99_out}
\end{figure}
\subsection{Boundary layer thickness}\label{sec:delta99R}
The growth of the boundary layer thickness $\delta_{99}(\hat{x})$ downstream of the rough-to-smooth transition is compared with $\delta_{99R}$, the thickness that a boundary layer developing on a homogeneously rough surface would have. $\delta_{99R}$ is computed by integrating the von K\'arm\'an momentum integral equation \citep{monty2016assessment} in the streamwise direction. The ratio $\delta_{99}(\hat{x})/\delta_{99R}(\hat{x})$ for all cases is presented in figure \ref{fig:delta99_out}. This ratio is very close to 1 (shown by the solid black line) in the vicinity of the roughness transition, suggesting that the growth rate of a rough-wall boundary layer thickness is initially sustained after the transition. Farther downstream, a consistent deviation from unity is observed, especially in case \texttt{Re07ks16} which has the longest fetch on the downstream smooth wall. A $\delta_{99}/\delta_{99R}$ ratio smaller than 1 indicates that the boundary layer thickness after a rough-to-smooth change is smaller than its rough-wall counterpart at the same fetch. A reference $\delta_{99}/\delta_{99R}$ curve can be further provided assuming that the entire boundary layer immediately adjusts to a smooth-wall mean velocity profile with matched $\delta_{99}$ right after the rough-to-smooth transition, which is also computed following the same method of \citet{monty2016assessment} and shown by coloured solid lines in figure \ref{fig:delta99_out}. This is referred to as the `immediate-change reference' in the following text. All data points of the present cases are above the immediate-change reference lines.

\subsection{Velocity scale}
Elliott's (\citeyear{Elliott1958}) model of the developing flow downstream of a rough-to-smooth transition assumes that above the IBL, $U_{\tau0}$ (the friction velocity over the upstream rough surface) is the correct velocity scale, which implies that the over-energised outer turbulence continues to reflect the upstream roughness velocity scale, even far downstream of the transition, and does not decay or relax in the downstream.

\begin{figure}
\centering
\setlength{\unitlength}{1cm}
\begin{picture}(13,5.1)
\put(0,0){\includegraphics[scale = 0.95]{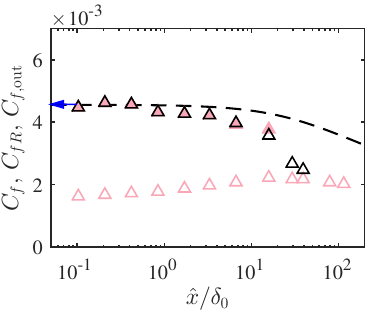}}
\put(6.5,0){\includegraphics[scale = 0.95]{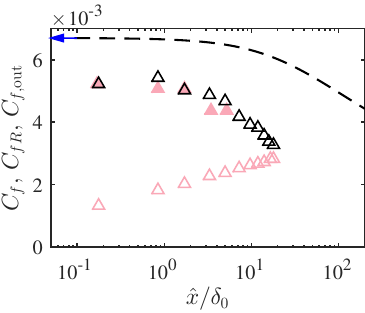}}
\put(0.1,4.8){(\textit{a})}
\put(6.6,4.8){(\textit{b})}
\end{picture}

\caption{$C_f$ obtained within the IBL and above the IBL via curve fitting for (\textit{a}) case \texttt{Re07ks16} and (\textit{b}) case grit adapted from \cite{Hanson2016}. Empty pink triangles in both figures are $C_f$ measured in the near-wall region, while empty black and filled pink triangles are $C_{f,\mathrm{out}}$ computed assuming an outer-layer similarity in the mean velocity deficit with $\delta_i$ estimated following the method of \cite{Antonia1971a} and the thresholding approach as detailed in \S\ref{sec:big_IBL}, respectively. The black dashed line is $C_{fR}$ of a homogeneous rough-wall boundary layer without the change of surface condition at $x_0$ computed by integrating the von K\'arm\'an momentum integral equation \citep{monty2016assessment}, and the blue arrow shows the skin-friction coefficient at the reference upstream rough-wall measurement.}
\label{fig:Cfo}
\end{figure}

\cite{Hanson2016} calculated the friction velocity $U_{\tau, \mathrm{out}}$ required to force the collapse of the mean velocity deficit on a smooth-wall reference above the internal layer. They reported that all $C_{f,\mathrm{out}}(\equiv 2U_{\tau, \mathrm{out}}^2/U_{\infty}^2)$ values are less than $C_{f0}$, the skin-friction coefficient on the upstream rough surface, and it decays in streamwise and approaches (though has not reached) the local smooth wall $C_f$ at the last measuring location at $\hat{x}/\delta_0 \approx 20$. Data from their original figure 19(\textit{a}) are reproduced here in figure \ref{fig:Cfo}(\textit{b}) for comparison. The $C_{f,\mathrm{out}}$ for case \texttt{Re07ks16} computed following the same procedure as in \cite{Hanson2016} is shown in figure \ref{fig:Cfo}(\textit{a}). Note that similar behaviours are observed in all other cases, and are not shown here for brevity. Two definitions of $\delta_i$ are tested: the classic one based on the $U$ versus $z^{1/2}$ profile originally proposed by \cite{Antonia1971a} and used in \cite{Hanson2016}, and the current definition based on the turbulence intensity profile as detailed in \S\ref{sec:big_IBL}. The resulting $C_{f,\mathrm{out}}$ with $\delta_i$ computed from these two definitions are shown in figure \ref{fig:Cfo}(\textit{a}) by empty black and solid pink triangles, respectively. In the present study, $\delta_i$ is defined as the upper limit of the (assumed) fluctuating rough/smooth interface, while the definition used by \cite{Hanson2016} appears to be closer to the averaged location of the (assumed) fluctuating interface. As $\delta_i$ grows and approaches the local $\delta_{99}$, the fitting range of the outer layer diminishes and eventually there are not enough data points to perform the regression, making the $C_{f,\mathrm{out}}$ trend in the limit of large $\hat{x}/\delta_0$ unattainable. There is little difference between the two $C_{f,\mathrm{out}}$ results except for the last few downstream locations, where the solid pink triangles (with $\delta_i$ defined based on the turbulence intensity profile) are absent due to the vanishing of the fitting range as $\delta_i$ approaches $\delta_{99}$. With Antonia \& Luxton's definition of $\delta_i$, a larger portion of the velocity profile which has in fact been affected by the new smooth-wall condition is included in the fit, therefore, a distinct decrease presents in $C_{f,\mathrm{out}}$, qualitatively similar to the near collapse in $C_{f, \mathrm{out}}$ and $C_f$ at their most downstream station as reported by \cite{Hanson2016}. 

The skin-friction coefficient for a homogeneously rough surface (with the same roughness as the upstream surface) $C_{fR}(\hat{x})$ is also included for comparison in figure \ref{fig:Cfo} (black dashed line), which is computed through the same streamwise evolution as that employed in \S\ref{sec:delta99R} to obtain $\delta_{99R}$\footnote{To predict the corresponding $C_{fR}(\hat{x})$ for Hanson and Ganapathisubramani's dataset, we use $\Pi_j = 0.71$ as reported for their homogeneous smooth-wall boundary layer. $k_s$ is calculated from the reported viscous-scaled roughness length $z_0^+$ of the rough surface. $\kappa = 0.41$ and $B = 5.0$ are used in the evolution to be consistent with their choice of the constants. The initial condition of the evolution is adjusted such that the resulting $C_{fR}(0)$ matches the reported skin-friction coefficient just upstream of the roughness transition.}. Just prior to the surface transition, the $C_{fR}(0)$ obtained from the measured upstream rough-wall profile is shown by a blue arrow on the logarithmic abscissa. Figure \ref{fig:Cfo} shows that $C_{f,\mathrm{out}}$ follows the predicted value for a homogeneously rough surface closely in the present dataset, while an immediate decrease in $C_{f,\mathrm{out}}$ compared to its value on the rough wall upstream is reported by \cite{Hanson2016}, which can be readily observed in (\textit{b}). In figure \ref{fig:Cfo}(\textit{b}), $C_{f,\mathrm{out}}$ at the first downstream location is lower than the rough-wall reference $C_{fR}(0)$ by $22\%$. Such a sudden change in the velocity scale in the outer layer challenges the traditional view that the effect of the new surface condition gradually modifies the interior of the flow through the growth of the IBL. Future work is required to further verify the existence of this behaviour and to understand the mechanisms behind it. Certainly, for the present measurements (figure \ref{fig:Cfo}\textit{a}), this effect is not observed.

\begin{figure}
\centering
\setlength{\unitlength}{1cm}
\begin{picture}(13,5.1)
\put(0,0){\includegraphics[scale = 0.95]{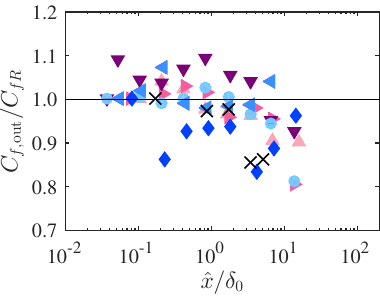}}
\put(6.5,0){\includegraphics[scale = 0.95]{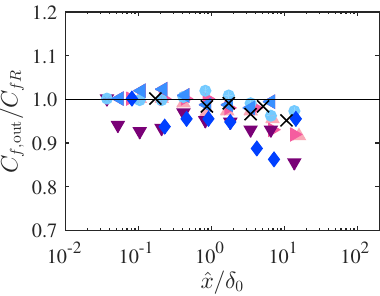}}
\put(0.1,4.8){(\textit{a})}
\put(6.6,4.8){(\textit{b})}
\end{picture}

\caption{$C_{f,\mathrm{out}}$ normalised by $C_{fR}$, the skin-friction coefficient for a homogeneously rough surface. $C_{f,\mathrm{out}}$ is calculated by assuming an outer-layer similarity in (\textit{a}) mean velocity deficit and  (\textit{b})  turbulence intensity profiles. The coloured filled symbols represent cases in the present study, and the black crosses are computed from the dataset of \cite{Hanson2016}. The solid black line is $C_{f,\mathrm{out}}/C_{fR} = 1$.}
\label{fig:CfoU_Uvar}
\end{figure}

We further examine the change of $C_{f,\mathrm{out}}$ with $\hat{x}$ for all cases. Figure \ref{fig:CfoU_Uvar} shows $C_{f,\mathrm{out}}(\hat{x})$ normalised by $C_{fR}(\hat{x})$ for all cases. $C_{f,\mathrm{out}}$ is computed by assuming an outer-layer similarity in the mean velocity deficit and turbulence intensity profiles in (\textit{a}) and (\textit{b}), respectively. If the velocity scale (i.e. rough-wall friction velocity) in the outer layer continues to develop as before without being affected by the rough-to-smooth transition, then $C_{f,\mathrm{out}}/C_{fR}$ should equal 1, as shown by the solid black line. Note that the focus here is on the trend of the streamwise evolution of $C_{f,\mathrm{out}}/C_{fR}$ in the streamwise direction rather than to assess how well the skin-friction coefficient predicted using the von K\'arm\'an momentum integral  equation matches the measurement on the rough wall. For this reason we divide all the downstream points by $C_{f,\mathrm{out}}/C_{fR}$ at the first downstream measuring station,  forcing the points immediately downstream of the transition in figure \ref{fig:CfoU_Uvar} to 1. This reduces the scatter of the data and facilitates the identification of an overall trend. The correction is usually less than $15\%$ in (\textit{a}) and $3\%$ in (\textit{b}), presumably indicative of the accuracy with which $C_{f,\mathrm{out}}$ can be determined by the two methods, and it does not show any dependence on $Re_{\tau 0}$ or $k_{s0}^+$.

Due to the uncertainty of the measurement and the fitting method, $C_{f,\mathrm{out}}$ computed from the mean velocity deficit (figure \ref{fig:CfoU_Uvar}\textit{a}) exhibits $\pm10\%$ scatter, while a less scattered trend is observed when the streamwise turbulence intensity profile is used (figure \ref{fig:CfoU_Uvar}\textit{b}). Overall, $C_{f,\mathrm{out}}/C_{fR}$ shows a slightly decreasing trend, and for all cases, it falls below 1 (shown by the solid black line) at $\hat{x}/\delta_0\approx 15$, the last streamwise location where $C_{f,\mathrm{out}}$ is attainable. For the homogeneously rough surface, $C_{fR}$ decreases in the streamwise direction as the boundary layer grows and the Reynolds number increases. The decreasing trend of the ratio $C_{f,\mathrm{out}}/C_{fR}$ suggests that the velocity scale in the outer layer decreases marginally more aggressively than the Reynolds number trend of $C_{fR}$, i.e. $C_{f,\mathrm{out}}/C_{fR}$ decreases slightly faster after a rough-to-smooth change than it would without the roughness transition. The physical interpretation is, the production of turbulent energy at the wall decreases after a rough-to-smooth change, and with a reduced supply of turbulent energy, the outer layer cannot sustain a rough-wall friction velocity that is no longer compatible with the new wall condition. Therefore, the velocity scale in the outer layer decreases more aggressively than the Reynolds number trend. The slow decay is presumably indicative of the time scale of the large-scale outer structures through which the boundary layer retains memory of upstream conditions.

Despite the small difference between $C_{f,\mathrm{out}}$ and $C_{fR}$, this further supports the conclusion in \S\ref{sec:big_mean} and \S\ref{sec:big_turb} that the velocity scale in the flow above the IBL is very close to the friction velocity of the rough wall upstream, and may in fact prove to be a reasonable modelling assumption in Elliott's approach.

It should be stressed that the determination of both $C_{f,\mathrm{out}}$ and $\delta_{99}$ suffer from a degree of uncertainty, however, we still observe a weak trend in both of them that they fall below the corresponding rough-wall value in the far field. The general picture is that when a turbulent boundary layer encounters a rough-to-smooth change, the change is barely felt in the outer-layer at first and the flow in the outer layer continues to evolve in streamwise as if the roughness transition was not there. Farther downstream, both $C_{f,\mathrm{out}}$ and $\delta_{99}$ begin to slowly adapt to the smooth-wall state, presumably as the over-energised large-scale structures in the outer layer, which retain the memory of the upstream conditions, decay and are replaced by events that reflect the new surface condition. A noticeable deviation from the rough-wall trend only appears after approximately $10\delta_0$ downstream of the transition.

\section{Procedure to apply the blending model}

In this section, we summarise the procedure to generate a velocity profile using the blending model described in \S\ref{sec:blending}. The upstream rough-wall velocity profile, the IBL thickness $\delta_i$, local friction velocity $U_{\tau2}$ and boundary layer thickness $\delta_{99}$ at a certain distance downstream of the rough-to-smooth change are required as input to the model. The steps are detailed as follows.

\begin{enumerate}
\renewcommand{\theenumi}{\roman{enumi}}
\item Generate the blending function $E(z^+)$ from (\ref{eq:blend_log}) \& (\ref{eq:blend_mu_sigma}). 
\item Generate the smooth-wall limit $U_S^+(z^+)$, i.e. a composite velocity profile \citep{Chauhan2009} with $Re_{\tau} = \delta_{99}U_{\tau 2}/\nu$.
\item Compute the rough-wall limit $U_R^{*+}(z_R^{*+})$ following (\ref{eq:blend_UR}) \& (\ref{eq:blend_zR}).
\item Obtain the final blending velocity profile at the given streamwise location through (\ref{eq:blend}).
\end{enumerate}


\bibliographystyle{jfm}
\bibliography{MAIN_BIB}
\end{document}